\documentclass[prl,reprint,superscriptaddress,amsmath,amssymb,longbibliography,aps]{revtex4-1}

\usepackage{graphicx}

\usepackage{mathrsfs}
\usepackage{mathbbol}
\usepackage{xcolor,booktabs}
\usepackage{overpic}
\usepackage{esint}
\usepackage{soul}

\newcommand{\ve}[1]{\ensuremath{\mbox{\boldmath$#1$}}}
\newcommand{\ma}[1]{\ensuremath{\mathbb{#1}}}

\usepackage{ragged2e}  
\usepackage[normalem]{ulem}
\usepackage{amsmath}

\usepackage{xcolor}

\begin{document}

\title{Torques on curved atmospheric fibres}

\author{F. Candelier}
\affiliation{Aix-Marseille Univ.,  CNRS, IUSTI, Marseille, France}
\author{K. Gustavsson}
\affiliation{Department of Physics, Gothenburg University, Gothenburg, SE-40530 Sweden}
\author{P. Sharma}
\affiliation{Max Planck Institute for Dynamics and Self-Organization, 37077 G\"ottingen, Germany}
\author{L. Sundberg}
\affiliation{Department of Physics, Gothenburg University, Gothenburg, SE-40530 Sweden}
\author{A. Pumir}
\affiliation{Laboratoire de Physique, ENS de Lyon, Universit\'e de Lyon 1 and CNRS, 69007 Lyon, France}
\affiliation{Max Planck Institute for Dynamics and Self-Organization, 37077 G\"ottingen, Germany}
\author{G. Bagheri}
\affiliation{Max Planck Institute for Dynamics and Self-Organization, 37077 G\"ottingen, Germany}
\author{B. Mehlig}
\affiliation{Department of Physics, Gothenburg University, Gothenburg, SE-40530 Sweden}

\date{\today}
\begin{abstract}
 Small particles are transported over long distances in the atmosphere, with significant environmental impact. The transport of symmetric particles is well understood, but atmospheric particles, such as curved microplastic fibres or ash particles, are generally asymmetric. This makes the description of their transport properties uncertain. Here,  we derive a model for how  planar curved fibres settle in quiescent air. The model  explains  that fluid-inertia torques may align 
such fibres at oblique angles with gravity as seen in recent laboratory experiments, and shows that inertial alignment is a general and thus important factor for the transport of atmospheric particles.
\end{abstract}


\maketitle

\emph{Introduction.}--  
Atmospheric transport  is a major dispersal pathway of microplastic particles~\cite{Allen_2021,Zhang_2020}, with negative environmental impact~\cite{Allen_2022,Prata_2020}, and
potentially affecting cloud-formation processes~\cite{Aeschlimann_2022,Revell_2021}.
Therefore it is important to develop reliable models for particle transport in the atmosphere~\cite{Allen_2022,Brahney:2021}. A challenge is 
that atmospheric particles,  such as curved microplastic fibres  \cite{cai2020origin,giurgiu2024full}, tend to be asymmetric. Theoretical models have focused on symmetric shapes~\cite{Xiao_2023,Gustavsson_2021}, which are now considered in the most advanced numerical models for atmospheric transport~\cite{Beckett:2022,Flexpart:2024}.
However, shape asymmetry may affect the transport of atmospheric fibers. This aspect is at best taken into account by empirical correlations, derived from experimental measurements~\cite{tatsii2024shape}.

Fluid-inertia torques may align fibres as they settle in the atmosphere. Particle-shape symmetry dictates how this happens: axisymmetric fibres with fore-aft symmetry for example  align so that their
broad side points down, as they settle in a quiescent fluid at moderate particle Reynolds numbers~\cite{cox1965steady,Khayat_Cox_1989,subramanian2005,Dabade_2015,Men17,roy2023orientation,Cabrera_2022,Jiang_2021}. 
In this case, only one angle matters (the angle between symmetry vector of the fibre and its centre-of-mass velocity).
This mechanism is important in the atmospheric sciences, as it
is thought to align symmetric ice platelets and columns  in weakly turbulent ice clouds \cite{Sassen_1980,Klett_1995,Breon_Dubrulle_2004,Noel_Sassen_2005,Gustavsson_2019,Gustavsson_2021}. 
Laboratory experiments in viscous cellular flow~\cite{Lopez_Guazzelli_2017}, in a turbulent water-oil mixture \cite{roy2023orientation},  for ensembles of fibres in quiescent and turbulent air~\cite{Newsom_Bruce_1994,newsom1998orientational,pierson2023inertial}, and for spheroids  in quiescent air \cite{bhowmick2024inertia} show that the inertial-torque models work quantitatively for axisymmetric particles with fore-aft symmetry. Moreover, breaking of fore-aft symmetry causes small axisymmetric particles to tilt as they settle \cite{Candelier_Mehlig_2016,Roy_2019,ravichandran2023orientation,maches2024settling}.
 \begin{figure}[t]
\begin{overpic}[width=0.75\columnwidth]{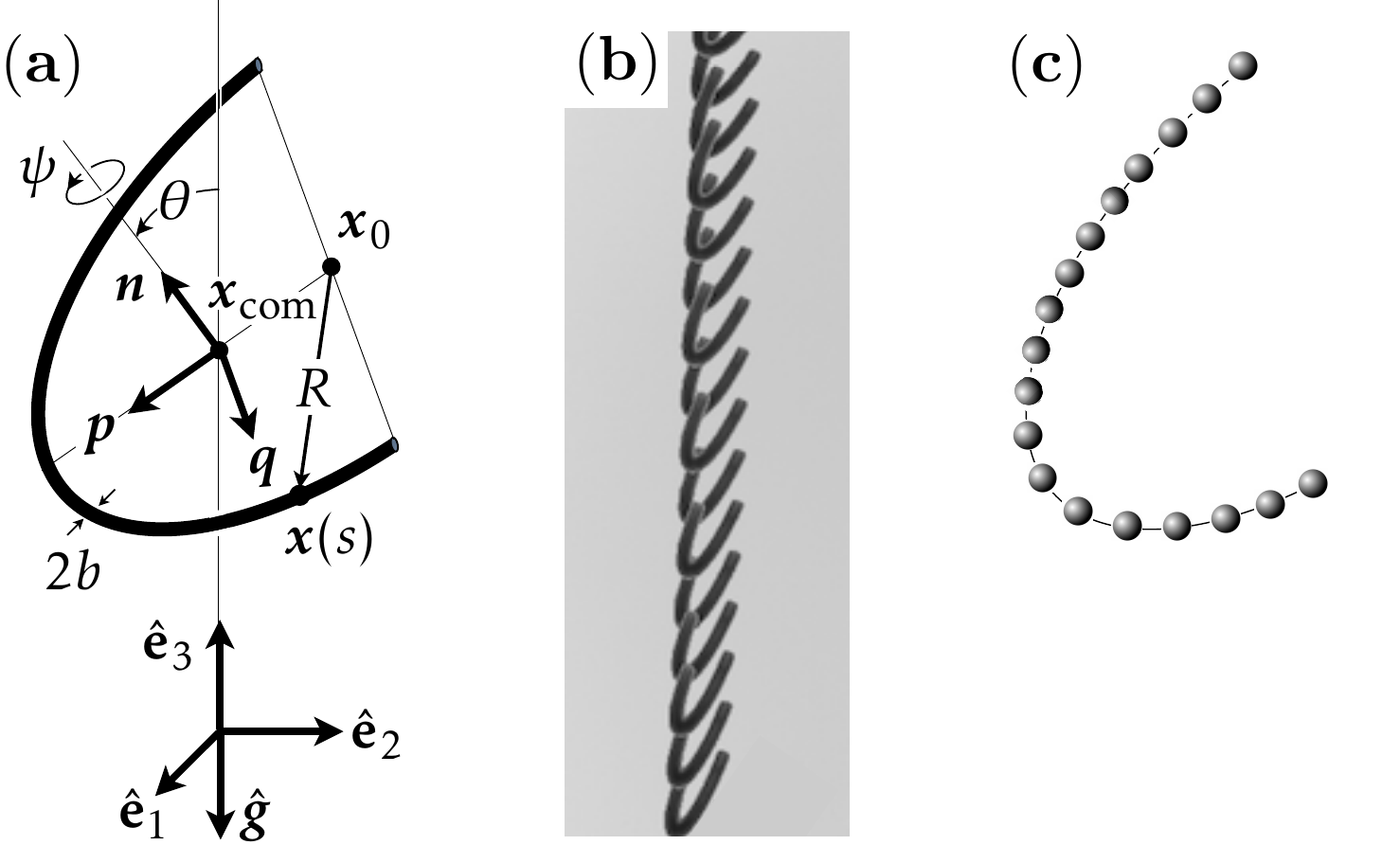}
\end{overpic}
\caption{\label{fig:coordinatesystem}  ({\bf a}) Particle geometry and definition of particle-fixed coordinate system $\ve p, \ve q, \ve n$. 
The angles $\theta$ and $\psi$ are Euler angles 
($Z_\phi Y_\theta Z_\psi$ convention, see supplemental material \cite{SM} for details). We refer to $\theta$ as the tilt angle, and $\psi$ as the spin angle. For $\phi\!=\!\theta\!=\!\psi\!=\!0$, the system $\ve p, \ve q, \ve n$ 
 aligns with the lab system $\hat{{\bf e}}_1,
 \ldots, \hat{{\bf e}}_3$. 
Also shown is the origin  $\ve x_0$ of the particle-fixed coordinate system, the centre-of-mass $\ve x_{\rm com}$, and the two radii of the fibre, $b$ and $R$. 
 ({\bf b}) Stacked video stills of microplastic fibre settling in air with an oblique tilt angle,
data from Ref.~\cite{tatsii2024shape}
. Gravity $\hat{\ve g}$ points down.
({\bf c})  Illustration of the model used in the Stokes limit, consisting of hydrodynamically interacting beads. }
\end{figure}

\begin{figure}
\centering
\begin{overpic}[width=\columnwidth]{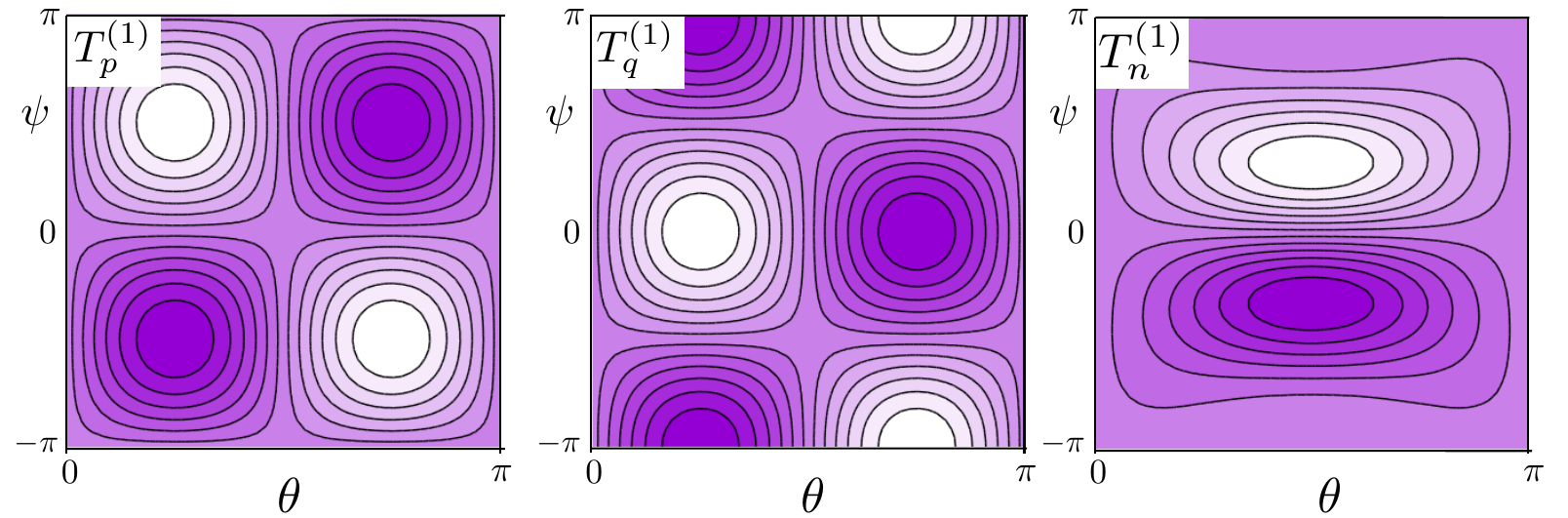}
\end{overpic}
\caption{\label{fig:half_circle_torques} {\em Inertial torques for the semi-circular fibre}.
Shown are the angular dependencies of  $T^{{(1)}}_p/T_p^{({1},{\rm max})}$, $T^{{(1)}}_q/T_q^{({1},{\rm max})}$, and $T^{{(1)}}_n/T_n^{({1},{\rm max})}$ 
of the inertial torque around the three axes of the particle-fixed coordinate system, obtained from  Eq.~(\ref{eq:torque_re}) for  $\hat{\ve v} = -\hat{\bf e}_3$ 
 ($\hat v_p = \sin\theta \cos\psi, \hat v_q =-\sin\theta\sin\psi$, and $\hat v_n = -\cos\theta$). Colourbar: from $-1$ (white) to $1$ (violet).
}
\end{figure}

Curved atmospheric fibres possess neither fore-aft symmetry nor are they axisymmetric. 
How do shape asymmetries of curved fibres affect inertial torques and thus the angular dynamics? To answer this question, we developed a model for the angular dynamics of curved microplastic fibres.
The model predicts how forces and torques depend on the fibre orientation, parametrised by its tilt angle $\theta$ and its spin angle $\psi$, see Fig.~\ref{fig:coordinatesystem}({\bf a}), as well as its shape and size. 
The model explains why current transport models for atmospheric fibres \cite{Xiao_2023,Beckett:2022,Flexpart:2024} work for  the component of the settling speed in the direction of gravity, but fail to describe
the sensitive shape-dependence of the 
 angular dynamics seen in the experiments \cite{tatsii2024shape}.
 In particular, fibres may align at oblique angles [Fig.~\ref{fig:coordinatesystem}({\bf b})], resulting for example in fibre transport perpendicular to gravity. 
 More generally, our results highlight how shape asymmetry affects the delicate effects of fluid inertia for typical atmospheric particles.

\emph{Model.}-- 
Newton's equations of motion read
\begin{subequations}
\label{eq:eom}
\begin{align}
\label{eq:force0}
\tfrac{\rm d}{{\rm d}t} \ve x_{\rm com} = \ve v \,,&\quad m\dot { \ve v }= \ve F_{\rm h} + \ve F_g\,,\\
\tfrac{\rm d}{{\rm d}t} \ve n= \ve\omega \wedge \ve n\,,&\quad
\tfrac{\rm d}{{\rm d}t} \ve p= \ve\omega \wedge \ve p\,,\quad \tfrac{\rm d}{{\rm d}t} (\ma J\ve \omega) = \ve T_{\rm h}\,.\label{eq:torque0}
\end{align}
\end{subequations}
with mass $m$, centre of mass $\ve x_{\rm com}$ and  inertia tensor $\ma J$ of the fibre, gravity force $\ve F_g=m\ve g$,
hydrodynamic force $\ve F_{\rm h}$ and torque $\ve T_{\rm h}$. 
The vectors $\ve p$, $\ve q$, and $\ve n$ denote the particle-fixed coordinate system  [Fig.~\ref{fig:coordinatesystem}({\bf a})]. 
We consider fibres in the form of circle segments with different radii of curvature, $R$. Examples are semi-circular fibres and quarter circles. 
The fibre diameter is denoted by $2b$, the contour length by $2a$. The aspect ratio  is $\kappa = a/b$. 

\emph{Stokes limit.}--
In a quiescent fluid, the Stokes force and torque on the particle  are~\cite{Kim_Karrila_1991}:
\begin{align}
\label{eq:ft}
\begin{bmatrix}\ve F_{\rm h}^{(0)}\\\ve T_{\rm h}^{(0)}\end{bmatrix}
=
-\mu \begin{bmatrix}\ma A & {\ma B}^{\sf T}  \\\ma B & \ma C  \end{bmatrix}
\begin{bmatrix} \ve v \\\ve \omega \end{bmatrix}\,.
\end{align}
Here $\ve v$ is the centre-of-mass velocity, $\ve \omega$ is the angular velocity around $\ve x_{\rm com}$, $\mu$ is the dynamic viscosity of the fluid, 
and $\ma A,\ma B, \ma C $ are resistance tensors. 
These tensors are constrained by particle-shape symmetries \cite{Happel_Brenner_1983,fries2017angular}.
Two reflection symmetries 
 cause all elements of $\ma B$ to vanish in the particle-fixed coordinate system, except $B_{qn}=\ve q\cdot\ma B\ve n$, and $B_{nq}$.
The tensors $\ma A$ and $\ma C$ are diagonal in this coordinate system.

In order to determine how
the fibres settle in the Stokes limit, we seek the steady states of Eqs.~(\ref{eq:eom}) with $\ve F_{\rm h}=\ve F_{\rm h}^{(0)}$ and $\ve T_{\rm h}=\ve T_{\rm h}^{(0)}$. This is accurate when the 
particle Reynolds number Re$_a=a v^*/\nu$ is very small, where $v^*$ is the steady-state settling speed of the fibre, and $\nu$ is the kinematic viscosity of air. 
Setting $\dot{\ve v}=\dot{\ve n}=\dot{\ve p}=\dot{\ve\omega}=0$  in Eqs.~(\ref{eq:eom}) yields two
steady states: 
$(\theta^\ast,\psi^\ast)=(\tfrac{\pi}{2},0)$ ($\ve p$ points down), and $(\tfrac{\pi}{2},\pi)$ ($\ve p$ points up).
Both steady states have $\ve\omega^*=0$ and $\ve v^*=(0,0,-v_g^*)$ with $v_g^*=mg/(\mu A_{pp})$.
Linear stability analysis  shows that 
the stability of the steady states depends on the signs of $B_{qn}$ and $B_{nq}$ (see supplemental material (SM)~\cite{SM} for details).
We computed the non-zero elements of the resistance tensors using the method of Durlofsky~\cite{durlofsky1987dynamic,collins2021lord,huseby2024helical}, 
modelling the fibres as chains of hydrodynamically  interacting beads with radius~$b$ [Fig.~\ref{fig:coordinatesystem}({\bf  c})].
This gives  $B_{qn}>0$ and $B_{nq}<0$ (Table~S1 in the SM~\cite{SM}). In this case,
$\psi^\ast=0$ is stable, while $\psi^\ast=\pi$ is unstable. 
 So the fibres settle with  $\ve n$ orthogonal to the settling velocity, and $\ve p$ points down.

\emph{Fluid-inertia torque.}-- At larger Re$_a$, the Stokes approximation fails because fluid-inertia torques affect the angular dynamics. 
We calculated the fluid-inertia torque for our curved fibres by expanding slender-body theory~\cite{Khayat_Cox_1989} to order Re$_a=a v^*/\nu$. In the (right-handed) particle-fixed coordinate system from Fig.~1({\bf a}) we find:
\begin{subequations} 
\label{eq:torque_re}
\begin{align}
 T_{p,M}^{(1)}/(2\pi \mu v a^2)& = 
\tfrac{ av}{\nu(\log \kappa)^2}  a_{p,M} {\hat v}_q {\hat v}_n\,,
\\
\label{eq:tqre} T_{q,M}^{(1)}/(2\pi \mu va^2) &=
 \tfrac{ av }{\nu(\log \kappa)^2}  a_{q,M} {\hat v}_p {\hat v}_n\,,\\
T_{n,M}^{(1)}/(2\pi \mu v a^2) &=
\tfrac{av }{\nu(\log \kappa)^2}
\big[a_{n,M}{\hat v}_q  {\hat v}_p\\
&+b_{n,M} {\hat v}_q + c_{n,M} \hat v_q 
\hat v_n^2+ d_{n,M} \hat v_q^3\big]\,.
\nonumber
\end{align}
 \end{subequations}
 Here $\hat v_p = \ve p \cdot \hat{\ve v}$  with $\hat {\ve v} = \ve v/v$, and so forth.
The parameter $M$ parameterises the central angle $2\pi/M$ of the circle segment
 ($M=2$ for a semi circle, $M=4$ for a quarter circle). 
The radius of curvature is $R=aM/\pi$. The coefficients $a_{p,M}, a_{q,M},\ldots$ 
 are given in 
 the SM \cite{SM}. 
Fig.~\ref{fig:half_circle_torques} shows the  angular dependence of the fluid-inertia torque~(\ref{eq:torque_re})
for a semi-circular fibre with $\hat{\ve v} = -\hat{\bf e}_3$.

Circle segments lack fore-aft symmetry and axisymmetry, 
but the remaining shape symmetry constrains how the components $T_p, T_q, T_n$ of the torque in the particle-fixed coordinate system depend on the components of the settling velocity $\ve v$. 
Covariance under coordinate change implies that  $T_p$  is even in $\hat v_p$, but odd in $\hat v_q$ and $\hat v_n$. 
Corresponding conclusions hold for $T_q$ and $T_n$.
Shape symmetry constrains the possible terms in Eq.~(\ref{eq:torque_re}) further, but allows for terms that do not occur there (the corresponding coefficients vanish). The reason may be that the slender-body approximation for $\ve T^{(1)}$ neglects hydrodynamic interactions between different parts of the fibre  to order $(\log\kappa)^{-2}$~\cite{keller1976slender,collins2021lord}. 

To understand how fluid inertia changes the angular
dynamics, we search for steady states using $\ve F_{\rm h}\!=\!\ve F^{(0)}_{\rm h}$ and $\ve T_{\rm h}\!=\!\ve T^{(0)}_{\rm h}\!+\!\ve T^{(1)}_{\rm h}$ in Eqs.~(\ref{eq:eom}), neglecting
 Re$_a$-corrections to the  force at first. Symmetry implies that  $v_q^\ast=0$ and $\psi^\ast=0$. Thus we seek solutions of
 \begin{subequations}
 \label{eqs:fp1} 
 \begin{align}
  \label{eq:F1}
  0&=    -\mu A_{pp} v_p + m g\sin\theta \,,\\
   \label{eq:F2}
  0 &= -\mu A_{nn} v_n - mg\cos\theta\,,\\
  0 &= -\mu B_{qn} v_n + 
  \tfrac{\mu a^3 }{\nu (\log\kappa)^2} 2\pi a_{q,M} v_p v_n\,.
  \label{eq:Tq2}
 \end{align}
 \end{subequations}
Eqs.~(\ref{eqs:fp1}) admit two steady states, 
 one with $v_n^\ast=0$, and the other one
 with $v_n^\ast\neq 0$.
 The first steady state is stable when $T_q^{(1)}$ is weak enough, and it is the same
as in the Stokes limit: 
 $\theta^*=\pi/2$,  $\psi^*_1=0$, $v_q^*=v_n^\ast=0$, and $v_p^*=v_g^\ast= mg/(\mu A_{pp})$.
This fixed point becomes unstable in a pitchfork bifurcation
as the fluid-inertia torque increases.
The bifurcation occurs when
\begin{equation}
{A'_{pp}B'_{qn}  (\log \kappa)^2}
    ={2\pi a_{q,M}\, \mathscr{R}\mathscr{V}}\,.
\end{equation}
Here we use non-dimensional variables and parameters:
$A'_{pp} = A_{pp}/a$, $B'_{qn} = B_{qn}/a^2$, 
 the particle-to-fluid mass-density ratio $\mathscr{R}=\rho_p/\rho_f$, and $\mathscr{V}=2\pi ab^2 g /\nu^2=\pi d_{\rm eq}^3g/(6\nu^2)$, a  non-dimensional volume~\cite{bhowmick2024inertia},
 and $d_{\rm eq}$ is the volume-equivalent diameter of the fibre. 
The product $\mathscr{RV}$ is proportional to Ga$^2$. The Galileo number Ga measures the ratio of
gravity to viscous forces.
For small particle Reynolds numbers,  $\mathscr{RV}\propto {\rm Re}_a$ up to a geometry-dependent factor. Thus larger $\mathscr{RV}$ implies larger fluid-inertia torques. Beyond the bifurcation -- for large enough $\mathscr{RV}$ -- two stable steady states form, with
stable orientation
\begin{equation}
\label{eq:thetaast}
\theta^\ast = 
     {\rm arcsin}\big( \tfrac{A'_{pp}B'_{qn}  (\log \kappa)^2}
    {2\pi a_{q,m}\, \mathscr{R}\mathscr{V}}\big)
\end{equation}
(and the equivalent $\pi-\theta^\ast$ due to symmetry).
We conclude: the plane containing the curved fibre in its steady state orientation tilts 
 when the argument of arcsin in Eq.~(\ref{eq:thetaast}) is
smaller than unity, so that
$\theta^\ast \neq \pi/2$. 
 Eq.~(\ref{eq:thetaast})  for $\theta^\ast$
is qualitatively similar to that
for axisymmetric particles with broken fore-aft symmetry \cite{Candelier_Mehlig_2016,Roy_2019,maches2024settling}. For the curved fibres, Eq.~(\ref{eq:thetaast}) is 
a consequence of broken symmetry under $\ve p \to -\ve p$.  
We note that fluid inertia does not change the value of $\psi^\ast$. 

\begin{table}[b]
\caption{\label{tab:exp_summary} 
Summary of experimental results for semi- and quarter-circular fibres. Parameters:  contour length $2a$ [mm], diameter $2b$ [mm], average settling speed $\langle v_g\rangle$ [m/s]
from Table~1 in Ref.~\cite{tatsii2024shape},  
experimental particle Reynolds number Re$_a \approx a \langle v_g\rangle /\nu$,
with
kinematic viscosity of air $\nu = 1.5 \times 10^{-5}\,$m$^2$/s, inertia parameter $\mathscr{RV}$ (see text). Steady-state angles $\psi^\ast$ [rad]  and $\theta^\ast$ [rad] [Fig.~\ref{fig:coordinatesystem}({\bf a})] extracted from the data \cite{tatsii2024shape} as  described in the SM~\cite{SM}. Number $N$ of 
experiments analysed.}
\mbox{}\\
\begin{tabular}{llllllllll}
 \hline\hline
         $\kappa$ &
         $2a$ &  $2b$  & $\langle v_g\rangle$ &${\rm Re}_a$  & $\mathscr{RV}$&\!\!\!\!~~~$\psi^\ast$ & $\theta^\ast$  & $N$ \\\hline\\[-4mm]
          \multicolumn{9}{c}{semi-circular fibres}\\
20& 2 & 0.1 &0.704& 47&660& --\footnote{\lq{}--\rq{} no steady state observed.}&--$^{\rm a}$ & 9\\
& 1 & 0.05& 0.288& 9.6 &83&0.013$\pm$0.024&1.21$\pm$0.1 & 8\\
50& 2 & 0.04&0.218 & 14.5 & 110&0&0.79 & 1\\
& 1 & 0.02 &0.086&2.9&13&0&1.54 & 1\\
100& 2 &0.02 &0.094&6.2& 26& 0&1.48$\pm$0.13 &5\\
  & 1 & 0.01&0.037& 1.24&3.3&{*}\footnote{\lq *\rq{}  fibre images blurred, orientations could not be determined. }&{*}$^{\rm b}$&0
  \\
    \multicolumn{9}{c}{quarter-circular fibres}\\
20& 2 & 0.1 &0.631&42& 660&--$^{\rm a}$ &--$^{\rm a}$ & 8\\
& 1 & 0.05&0.265&8.8& 83&-0.03$\pm$0.01 &1.34$\pm$0.13& 10\\
& 0.5& 0.025&0.096&1.6&10& -0.017&1.414&1\\
50 & 2 & 0.04&0.209&14.0 &110&-0.003$\pm$0.017& 1.48$\pm$0.06&11\\
& 1 & 0.02 &0.074&2.5&13&0&1.43$\pm$0.07&2\\
100& 2 &0.02&0.080&5.3 &26& 0&1.379&1\\
     & 1 & 0.01 &0.026&0.88&3.3&{*}$^{\rm b}$&{*}$^{\rm b}$&0\\
 \hline\hline\end{tabular}
\end{table}
\emph{Experimental data.}--  
Using the G\"ottingen
turret~\cite{bhowmick2024inertia}, \citet{tatsii2024shape} measured the settling speeds of curved fibres (quarter and  semi circles) with different lengths $L$ and aspect ratios $\kappa$, and mass density $\varrho_p=1200\,$kg/m$^3$, comparable 
to that of common microplastic fibres in the atmosphere. Here we extracted the angular dynamics from the video recordings 
(see SM~\cite{SM} for details). 
The results  (Table~\ref{tab:exp_summary}) are qualitatively consistent with the theory outlined above, except for the fibres with  the largest fluid-inertia parameters $\mathscr{RV}$ which do not align. As predicted by the theory, 
all other fibres  settle with 
steady orientations with $\psi^\ast=0$.
For quarter circles, the steady tilt angle is  $\theta^\ast \approx \tfrac{\pi}{2}$, so $\mathscr{RV}$ is below the bifurcation analysed above.
Semi-circular fibres, by contrast, settle with 
a $\mathscr{RV}$-dependent tilt angle. As predicted by the theory,  $\theta^\ast \approx \tfrac{\pi}{2} $
for small $\mathscr{RV}$, and $\theta^\ast$ decreases as $\mathscr{RV}$ increases. 

\begin{figure}
\centering
\begin{overpic}[width=\columnwidth]{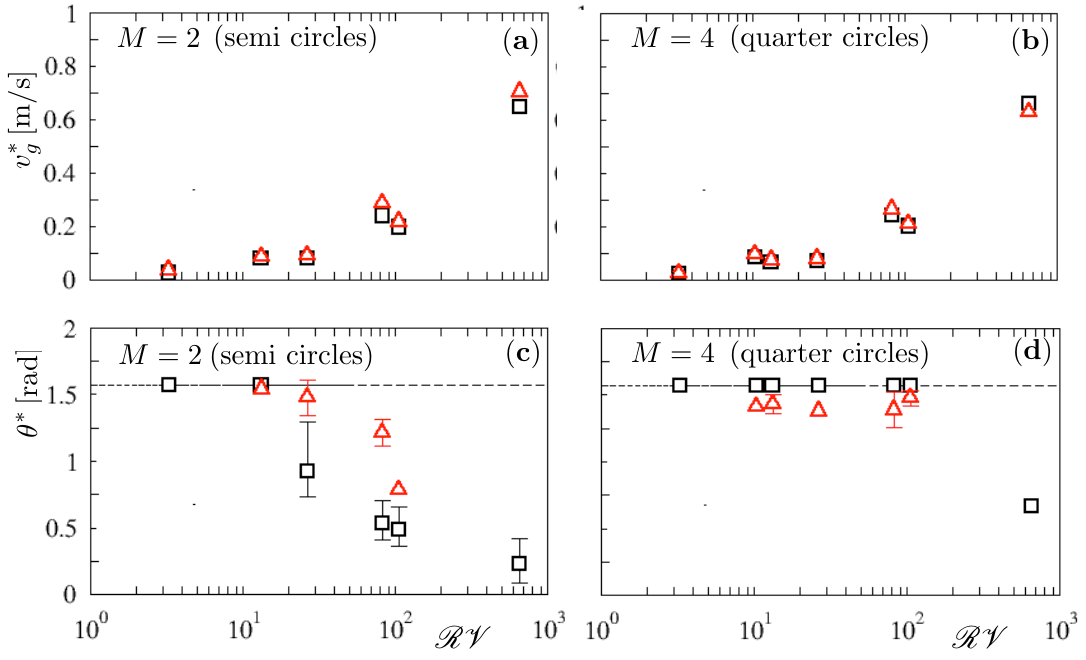}\end{overpic}
\caption{\label{fig:comparison} {\em Comparison between theory and experiment}
(Table~\ref{tab:exp_summary}).
({\bf a})~Full slender-body theory (see text) for 
$v_g^\ast$ 
for semi-circular fibres versus  
the inertia parameter $\mathscr{RV}$ ($\Box$); experimental $\langle v_g\rangle$ (red $\triangle$). ({\bf b}) same, but for quarter circles. 
({\bf c}) Theory for steady-state tilt angle $\theta^\ast$
for semi-circular fibres  ($\Box$). Estimated theoretical uncertainties (see text) are shown when larger than the symbol size.
Experiment (red $\triangle$) with errors from Table~\ref{tab:exp_summary}.  Dashed line corresponds to $\theta^\ast = \tfrac{\pi}{2}$.
({\bf d})~same, but for quarter circles (theoretical uncertainty not shown).}
\end{figure}

For a quantitative comparison, the model must be improved to 
 account for the fact that Re$_a$ is not small in the experiment, so that
 the small-Re$_a$ theory 
 (\ref{eq:torque_re}) does not directly apply. Moreover, we expect that fluid-inertia corrections
 to the force -- neglected above -- may be significant. To improve the model, 
 we  computed the fluid-inertia corrections to torque and force in the full slender-body theory \cite{Khayat_Cox_1989} by numerical integration of 
  Eqs.~(S13), (S14), and (S19) in the SM \cite{SM}. Then we 
 searched
for steady-state values $\ve v^\ast$ and  $\theta^\ast$ numerically,
for the fibres listed in Table~\ref{tab:exp_summary}.
The results of these calculations are summarised in Figure~\ref{fig:comparison}. Panels ({\bf a}) and ({\bf b})  compare
the experimental  $\langle v_g\rangle$ (Fig.~2a in Ref.~\cite{roy2023orientation}) for the fibres from Table~\ref{tab:exp_summary}
with the full theory for the gravity component $v_g^\ast$ of the steady settling velocity, as functions of  the fluid-inertia parameter $\mathscr{RV}$. We observe excellent agreement.

Panel ({\bf c}) compares the theory for $\theta^\ast$ with the experiments for semi-circular fibres. 
For small $\mathscr{RV}$, the Stokes torque dominates, yielding stable settling $\theta^\ast = \tfrac{\pi}{2}$. For larger $\mathscr{RV}$, the fibres settle at oblique angles with gravity. Compared with the experiment, theory gives smaller values of $\theta^\ast$ in the transition region between small and large $\mathscr{RV}$. Here the steady-state tilt angle is sensitive to the precise form of the hydrodynamic torque,
as evident from the estimate of the theoretical uncertainty shown in panel ({\bf c}) (see discussion below). 
Panel ({\bf d}) shows that the theory works well for the quarter circles too. 
For $\mathscr{RV}=660$, quarter and semi circles continue to tumble in the experiments. Our model  indicates that this may be explained by a second bifurcation around  $\mathscr{RV}=660$  that results in either
a long transient towards, or a change of stability of  $\theta^\ast$.
 
\emph{Discussion.}-- 
Our model shows
that the component $v_g^\ast$ 
of the settling velocity
is almost independent of the non-dimensional radius of curvature $R/a$, and only weakly dependent on its aspect ratio $\kappa$, in agreement with the experiment [Fig.~\ref{fig:comparison}({\bf a},{\bf b}) and Table~\ref{tab:exp_summary}]. 
For slender fibres, the weak $\kappa$-dependence is explained because the aspect ratio enters only as $\log\kappa$ in the model (at fixed $\mathscr{V}$).
 The weak dependence on $R/a$ is explained by the fact that the elements of $\ma A$ are similar for quarter  and semi circles, and that the $M$-dependence of the fluid-inertia force compensates for this dependence, in part. 
In summary, our model shows that the variation of $v_g^\ast$ for different fibres is mainly determined  by the parameter 
$\mathscr{RV}$. We conclude: this is the most important parameter for 
empirical settling models for slender fibres (but note that $g_g^\ast$ for a slender fibre differs from that of a sphere of equivalent volume \cite{bhowmick2024inertia,tatsii2024shape,Xiao_2023}).

Regarding the angular dynamics,
the conclusions are quite different.
  The experiments show that curved fibres settle at a steady
 tilt angle $\theta^\ast$ that depends sensitively on 
 the  non-dimensional radius of curvature $R/a$, as well as on $\mathscr{RV}$. In particular,  quarter  and semi circles settle quite differently
[Fig.~\ref{fig:comparison}({\bf c}),({\bf d})] for the same size and value of 
$\mathscr{RV}$. 
Our theory shows that this difference is a consequence of 
 the bifurcation.
The Stokes torque decreases more slowly than the fluid-inertia torque as one straightens the fibre by increasing $R/a$, 
shifting the location of the bifurcation for quarter circles to larger values of $\mathscr{RV}$. Beyond the bifurcation, the model predicts significant motion perpendicular to gravity that is not explained
by existing empirical models. 
We find $v_\perp/v_g\approx 0.13$ to $0.15$ for the
first three data points after the bifurcation in Fig.~\ref{fig:comparison}({\bf c}).
In conclusion, atmospheric transport models for slender fibres must account not only for $\mathscr{RV}$, but also for the radius of curvature. 
 
 The theory also explains why fluid inertia does not change 
the spin angle, it remains 
  $\psi^\ast=0$ as in the Stokes limit. This follows 
from the shape symmetry of the fibre. But note that
imperfections (shape asymmetries or density inhomogeneities) may  break this symmetry, causing the steady-state angle $\psi^\ast$ to change, distorting the patterns in Fig.~\ref{fig:half_circle_torques}. Note also that 
 non-planar fibres may continue to rotate in the Stokes limit \cite{gonzalez2004dynamics}, like chiral ribbons~\cite{huseby2024helical} and taco-shaped disks \cite{miara2024dynamics}. How fluid inertia affects this dynamics is not known. 
  
Overall, Fig.~\ref{fig:comparison} shows  good agreement between model and experiment, except near the bifurcation, 
where the tilt angle is sensitive to the details of the hydrodynamic torque. We stress that the model does not contain any fitting parameters, unlike that in e.g. Ref.~\cite{bhowmick2024inertia}.

We now discuss possible shortcomings of the full model, and the deviations between model and experiment for $\theta^\ast$ near the bifurcation. For straight cylinders, \citet{fintzi2023inertial} assessed the accuracy of slender-body theory by DNS, comparing to Ref.~\cite{Khayat_Cox_1989}, and to higher-order calculations \cite{khair2018higher}.
They  find that the theory of \citet{Khayat_Cox_1989} fails to quantitatively account for 
contributions to force and torque  from the lateral surfaces (end caps) of the cylinder for  Re$_a> 1$ and $\kappa < 30$. 
The upstream end cap causes larger deviations than the downstream one. Since the end caps of our steadily settling fibres tend to lie downstream, we expect slender-body theory to  work somewhat better for that orientation.
Near the bifurcation, $\theta^\ast$ depends sensitively on forces and torques, therefore it is hard to say whether these shortcomings of slender-body theory increase or decrease $\theta^\ast$. For this reason, 
we include an estimate of the theoretical uncertainty in Fig.~\ref{fig:comparison}({\bf c}), showing by how much $\theta^\ast$ changes when increasing/decreasing fluid-inertia torque and force 
by $\pm 10$\%. 
Slender-body theory for the Stokes torque requires even larger values of $\kappa$ \cite{keller1976slender}. We avoided the latter problem by computing the Stokes torque using the method of~Ref.~\cite{durlofsky1987dynamic}.

In some cases, the experiments show that steady-state alignment is approached in an oscillatory fashion, for example as $\psi(t) \sim  \cos(\omega t+\varphi) \exp(-\lambda t)$ (Fig.~S3 in the SM \cite{SM}). This is qualitatively explained by  linear stability theory in the Stokes limit.  Fluid inertia, however,  is expected to have a significant effect on the transient. This remains to be worked out.

For the largest values of $\mathscr{RV}$, the experiments show unsteady angular dynamics (no steady state  within $\sim 30$\,s, Fig.~S2 in the SM \cite{SM}). This is analogous to the DNS results of \citet{auguste2013falling} and  the experiments of  \citet{coletti2023} for disks settling at Re $\sim 100$, where periodic angular dynamics can coexist with steady alignment. Our model may qualitatively explain such behaviour, because our numerics indicates 
that there is a second bifurcation near $\mathscr{RV}=660$ where oblique settling becomes unstable. However, the DNS results for straight fibres of~\citet{fintzi2023inertial} indicate that our model is at best qualitatively correct for  $\mathscr{RV}=660$.

\emph{Conclusions.}--
Microplastic fibres in the atmosphere settle rapidly because their mass density is about 1000 times higher than that of air. Particle Reynolds numbers based on the fibre half-length 
 are of order 10 for 1mm fibres. Since such fibres tend to be curved, we 
analysed through model calculations how curvature affects the settling dynamics of curved fibres. 
 Our model explains why curved fibres may align at oblique angles, depending intricately on their shapes, causing transport perpendicular to the settling direction.
The model exemplifies how shape-symmetry breaking affects the angular dynamics of atmospheric particles. This is important because such
 particles are often asymmetric, just like the curved fibres discussed here, 
 atmospheric ice-crystals which usually have imperfections that break shape symmetries \cite{Heymsfield_1973}, and volcanic-ash particles. For atmospheric particles with highly irregular shapes 
 it may be of interest to generalise the present model to random aggregates. 

The angular dynamics of  symmetric particles settling in turbulent air has recently been studied in detail \cite{Gustavsson_2021,tinklenberg2024turbulence,bhowmick2024inertia}. 
 Asymmetric particles can exhibit intricate dynamics in the Stokes limit \cite{giurgiu2024full,tozzi2011settling}. In turbulent air, we expect that particle and fluid inertia change the dynamics more than in quiescent air, so the effects discussed here must be accounted for in transport models for atmospheric particles. To conclude, we stress that the conceptually simple case of curved fibers considered here is only  one example of the many asymmetric shapes found in the atmosphere, and that the present study uncovers essential physics relevant for  a broad class of atmospheric particles.

\begin{acknowledgments}
This work was supported by Vetenskapsr\aa{}det under grant nos. 2021-4452 (BM,LS), 2018-03974 and 2023-03617 (KG).  FC thanks J. Gissinger for testing the implementation of the bead model. We thank E. Bodenschatz for support and discussions.
\end{acknowledgments}


%
\end{document}


\title{Supplemental Material:\\``{Torques on curved atmospheric fibres}"}

\author{F. Candelier}
\affiliation{Aix-Marseille Univ.,  CNRS, IUSTI, Marseille, France}
\author{K. Gustavsson}
\affiliation{Department of Physics, Gothenburg University, Gothenburg, SE-40530 Sweden}
\author{P. Sharma}
\affiliation{Max Planck Institute for Dynamics and Self-Organization, 37077 G\"ottingen, Germany}
\author{L. Sundberg}
\affiliation{Department of Physics, Gothenburg University, Gothenburg, SE-40530 Sweden}
\author{A. Pumir}
\affiliation{Laboratoire de Physique, ENS de Lyon, Universit\'e de Lyon 1 and CNRS, 69007 Lyon, France}
\affiliation{Max Planck Institute for Dynamics and Self-Organization, 37077 G\"ottingen, Germany}
\author{G. Bagheri}
\affiliation{Max Planck Institute for Dynamics and Self-Organization, 37077 G\"ottingen, Germany}
\author{B. Mehlig}
\affiliation{Department of Physics, Gothenburg University, Gothenburg, SE-40530 Sweden}

\date{\today}
\begin{abstract}
In this document we summarise details regarding the theory (Sections I-V), the method of analysing the video recordings (Section VI), and the results obtained
from this analysis (Section VII). This document contains 13 supplemental figures and
one supplemental table.
\end{abstract}


\maketitle

\newpage
\setcounter{figure}{0}
\setcounter{table}{0}
\makeatletter
\renewcommand{\thefigure}{S\arabic{figure}}
\renewcommand{\thetable}{S\arabic{table}}
\renewcommand{\theequation}{S\arabic{equation}}
\renewcommand{\thesection}{S\arabic{section}}

\onecolumngrid

\section{Definition of Euler angles}
\label{app:euler}

We denote the orthogonal basis  of the lab-coordinate system by
$\hat{\bf e}_\alpha$, and parameterise the orientation of the body-coordinate system $\ve p$, $\ve q$, $\ve n$ using three angles $\theta, \phi, \psi$
as follows [see Fig.~1({\bf a}) in the main text]
\begin{align}
\nonumber
\ve n&=
\begin{bmatrix}
c_\phi s_\theta\cr
s_\phi s_\theta\cr
c_\theta
\end{bmatrix}
\,,\hspace{0.3cm}
\ve q=
c_\psi
\begin{bmatrix}
-s_\phi\cr
c_\phi\cr
0
\end{bmatrix}
{+}s_\psi\,\ve n\wedge\begin{bmatrix}
-s_\phi\cr
c_\phi\cr
0
\end{bmatrix}\,,\\
\ve p&=\ve q \wedge \ve n\,.
\label{eq:coordinate system}
\end{align}
where $s_\theta = \sin\theta, c_\phi = \cos\phi$, and so forth.
Here  $\ve p$ and $\ve q$ are orthogonal to $\ve n$,  such that $q_z=0$ when $\psi=0$, and $\ve p$ is chosen to form a right-handed coordinate system with $\ve q$ and $\ve n$, see
Fig.~1({\bf a}) in the main text.

This parameterisation is equivalent to 
Euler angles using the 
$Z_\phi Y_\theta Z_\psi$ convention.
Assume that the particle-fixed coordinate system aligns as follows, $\ve p$ with $\hat{\bf e}_1$, $\ve q$ with $\hat{\bf e}_2$, and $\ve n$
with $\hat{\bf e}_3$. First, rotate
the particle-fixed coordinate system 
around $\hat{\bf e}_3$ by~$\phi$, so that the image $\ve p'$ of $\hat{\bf e}_1$ rotates in the $\hat{\bf e}_1$-$\hat{\bf e}_2$-plane. 
The new rotated coordinate system is called $\ve p', \ve q', \ve n'=\ve n$. Next, rotate this coordinate system
 around $\ve q'$  by~$\theta$, the polar angle. The new coordinate system is called 
$\ve p^{\prime\prime}, \ve q^{\prime\prime}, \ve n^{\prime\prime}$.
The third step is to rotate this coordinate system around $\ve n^{\prime\prime}$ by $\psi$. All rotations are counter-clockwise. This yields the rotation matrix for the basis vectors 
\begin{equation}
\label{eq:R2}
\ma R = \begin{bmatrix}
c_\phi c_\theta c_\psi -s_\phi s_\psi & -s_\phi c_\psi -c_\phi c_\theta {s}_\psi&c_\phi s_\theta\\
c_\phi s_\psi + s_\phi c_\theta c_\psi & c_\phi c_\psi -s_\phi c_\theta s_\psi & s_\phi s_\theta\\
-s_\theta c_\psi & s_\theta s_\psi & c_\theta
\end{bmatrix}\,.
\end{equation}
The columns of $\ma R$ are $\ve p$, $\ve q$, and $\ve n$. Components $v_i$ of a vector $\ve v$ in the lab frame map to the particle-fixed coordinate system as
\begin{align}
    \begin{bmatrix} {v}_p\\{v}_q\\{v}_n\end{bmatrix}
    = \ma R^{\sf T} \begin{bmatrix} {v}_1\\{v}_2\\{v}_3\end{bmatrix}\,.
\end{align}
 To make the Euler angles  unique, one needs to restrict their ranges, because $[\phi,\theta,\psi]$ and $[\phi',\theta',\psi']=[\phi+\pi,-\theta,\psi+\pi]$ (any angle modulo $\pm 2\pi$) describe the same particle orientation. 
For our planar fibre, the multiplicity of equivalent triplets is higher than that,  because the fibre has two reflection symmetries. 
Applying both reflections corresponds 
to the matrix
\begin{equation}
\label{eq:S}
\ma S = \begin{bmatrix} 1 &&\\&-1& \\ && -1\end{bmatrix}
\end{equation}
in the particle-fixed coordinate system. 
The rotation matrices corresponding to the two triplets $[\phi,\theta,\psi]$ and $[\phi',\theta',\psi']=[-\phi,-\theta+\pi,\psi+\pi]$ (any angle modulo $\pm 2\pi$) are
related by
\begin{equation}
\label{eq:SR}
\ma S \ma R' =\ma R \,.
\end{equation}
Therefore the two triplets describe indistinguishable particle orientations. 
In conclusion, 
the four triplets
$[\phi,\theta,\psi]$, $[\phi+\pi,-\theta,\psi+\pi]$, $[-\phi, -\theta+\pi,\psi+\pi]$, and $[-\phi+\pi,\theta+\pi,\psi]$ (any angle modulo $\pm 2\pi$)
are equivalent because they yield indistinguishable particle orientations. We discuss in Section \ref{app:data} how the angles were restricted to extract unique values of $\theta$ and $\psi$ from the experiments.

\begin{table*}
    \caption{\label{tab:coeffs}Stokes resistance coefficients
    in the Stokes limit obtained using the method of \citet{durlofsky1987dynamic} (see text). }
     \begin{tabular}{ccccccccccc}
 \hline\hline
 $2a$ [mm] &  $2b$ [$\mu$m] & $A_{pp}$ [mm] & $A_{qq}$ [mm] & $A_{nn}$ [mm]& $B_{qn}$ [mm$^2$]& $B_{nq}$ [mm$^2$]
& $C_{pp}$ [mm$^3$]& $C_{qq}$ [mm$^3$]& $C_{nn}$   [mm$^3$]      
\\\hline\\[-2.5mm]
          \multicolumn{10}{c}{semi-circular fibres}\\[1mm]
2 & 100 & 5.390& 5.102 & 6.321  & 0.1889 & -0.6331 & 2.027  & 0.5619   & 2.015\\
1 & 50& 2.695&2.551&3.161&0.0472&-0.1583&0.2533&0.0702&0.2519\\
2 & 40& 4.215 &4.017& 5.060 & 0.1194 &  -0.5150 & 1.439& 0.3446 &1.354\\
1 &20& 2.107 & 2.009 & 2.531 & 0.02985 & -0.1288 & 0.1799 & 0.04307 & 0.1692\\
2 & 20& 3.624  & 3.471 & 4.408 & 0.08649  & -0.4485 & 1.186 & 0.2676 & 1.083\\
1 & 10&  1.812 & 1.736 & 2.204 & 0.02162 & -0.1121 & 0.1482 & 0.03345 & 0.1354\\[1mm]
    \multicolumn{10}{c}{quarter-circular fibres}\\[1mm]
 2 & 100& 6.104 & 4.677& 6.423&0.1209 & -0.6035 & 3.104&0.2242 &3.056 \\
 1 &  50&  3.052 &2.338 & 3.212  &  0.03021  & -0.1509  & 0.3880 &0.02802 &0.38204 \\
0.5&  25&  1.526 &  1.169& 1.606 &0.0075 &-0.0377 & 0.04850&  0.00350 & 0.04776\\
 2 &  40 & 4.831  & 3.579 &  5.125  & 0.07493. &  -0.4963 & 2.159  & 0.1207 &    2.094\\
 1 &  20& 2.416 & 1.790 & 2.562 & 0.01873 &  -0.12407 & 0.2699 & 0.01509 & 0.26171\\
 2 &  20& 4.183 & 3.042 & 4.457 & 0.05387 & -0.4373 & 1.762 & 0.0904 & 1.695  \\
 1 &  10& 2.091 & 1.521 & 2.228 & 0.01347 & -0.1093 & 0.2203 & 0.01130 & 0.2118\\
    \hline\hline\end{tabular}
\end{table*}

\section{Stokes resistance coefficients}
\label{app:durlofsky}
Table~\ref{tab:coeffs} summarises
the numerical values for the Stokes resistance coefficients for the fibres from Table~1 in the main text. The results
were obtained by modeling the fibre as a chain of hydrodynamically interacting beads~\cite{durlofsky1987dynamic,collins2021lord,huseby2024helical}.
We use $N+1$ beads of radius $b$ to represent the fibre [Fig.~1({\rm c}) in the main text], the distance between the bead centres is thus $a/N$. We determine this distance by computing the resistance tensor $\ma A$ for a slender straight wire and comparing with Eqs.~(19,20) of Ref.~\cite{keller1976slender}, an expansion to order $(\log\kappa)^{-3}$. The smallest difference is observed for $a/N=2.3b$. In this case the relative errors are 1.7 \% for $A_\parallel$
and 2.9 \% for $A_\perp$ for a chain with $300$ beads ($\kappa \approx 344$) \cite{gissinger2023dynamique}. Here $A_\parallel$ (resp. $A_\perp$) is the resistance coefficient when the fibre moves with a velocity collinear (resp. perpendicular) to its axis. 
\citet{kharrouba2021flow} computed the resistance coefficients
to order $(\log\kappa)^{-4}$. For the chain with $300$ beads, the fourth-order correction is negligible.

\section{Linear stability analysis in the Stokes limit}
\label{app:stability}
In the Stokes limit and in a quiescent fluid, Eqs.~(1) and (2) in the main text admit two fixed points, $\ve \omega^\ast=0$, $v^*_q=v^*_n=0$, $v_p^*={m} g \sigma_\psi/(\mu A_{pp})$, $\theta^\ast=\pi/2$, and $\psi^\ast=(1-\sigma_\psi)\pi/2$, where $\sigma_\psi=\pm 1$. 
Close to the fixed points, we expand the dynamics as $v_p=v_p^*+\delta v_p$,  and so forth. We obtain three decoupled systems of differential equations. 
The first system,
\begin{subequations}
\begin{align}
\tfrac{{\rm d}}{{\rm d}t}\delta v_p&=-\tfrac{A_{pp}\mu}{{m}}\delta v_p\,,\\
\tfrac{{\rm d}}{{\rm d}t}\delta\phi&=-\sigma_\psi\delta\omega_p\,,\\
\tfrac{{\rm d}}{{\rm d}t}\delta\omega_p&=-\tfrac{C_{pp}\mu}{J_{pp}}\delta\omega_p\,,
\end{align}
\end{subequations}
describes how $\delta v_p$ and $\delta \omega_p$ decay exponentially towards zero. 
Here $J_{pp}$ (and $J_{qq}, J_{nn}$) are the non-zero elements of the particle-inertia tensor $\ma J$ in the particle-fixed basis. 
The angle $\phi$ goes to an undetermined constant $\phi^\ast$ (a consequence of rotational symmetry around $\ve g$).
The second system,
\begin{subequations}
\label{eq:subsystem_psi}
\begin{align}
\tfrac{{\rm d}}{{\rm d}t}\delta v_q&=-\tfrac{A_{qq}\mu}{{m}}\delta v_q-\big[\tfrac{B_{nq}\mu}{{m}}+\tfrac{{m}g\sigma_\psi}{A_{pp}\mu}\big]\delta\omega_n - g\sigma_\psi \delta\psi \,,\\
\tfrac{{\rm d}}{{\rm d}t}\delta\psi&=\delta\omega_n \,,\\
\tfrac{{\rm d}}{{\rm d}t}\delta\omega_n&=-\frac{\mu}{J_{nn}}\big[C_{nn}\delta\omega_n+B_{nq}\delta v_q\big]\,,
\end{align}
\end{subequations}
has three stability exponents $\lambda_1,\ldots,\lambda_3$, obtained by solving a third-order equation.
In the underdamped limit of large rotational inertia $J_{nn}$ we obtain 
\begin{subequations}
\label{eq:system2}
\begin{align}
\lambda_{1,2}&\sim\tfrac{1}{2A_{qq}{J_{nn}}}\big[\sigma_{\psi}\tfrac{B_{nq}g{m}^2}{\mu}(\tfrac{1}{A_{pp}}-\tfrac{1}{A_{qq}})
-\mu(A_{qq}C_{nn}-B_{nq}^2)\big] \pm{\rm i}\sqrt{-{\sigma_\psi}\tfrac{B_{nq} {m}g}{A_{qq} J_{nn}}}\,,\\
\lambda_3&\sim -\frac{A_{qq}\mu}{{m}}\,.
\end{align}
\end{subequations} 
Since $A_{qq}C_{nn}-B_{nq}^2>0$, the dynamics damps to zero in the absence of gravity.
With gravity, the fixed point with $\sigma_\psi=1$, $\psi^*=0$, is stable. This follows because 
our particles have $A_{pp} > A_{qq}$ and $B_{nq}<0$. 
In this underdamped limit, the variables
$v_q$, $\psi$, and $\omega_q$ exhibit transient oscillations.
Their
angular velocity can be read off from the imaginary part of 
$\lambda_{1,2}$. 
Consider instead the overdamped limit ($m$ and $J_{nn}$ small). In this limit, one observes the same steady state, but the transient  does not exhibit oscillations. 
The third system,
\begin{subequations}
\begin{align}
\tfrac{{\rm d}}{{\rm d}t}\delta v_n&\!=\!-\tfrac{A_{nn}\mu}{{m}}\delta v_n\!-\![\tfrac{B_{qn}\mu}{{m}}\!-\!\tfrac{{m}g\sigma_\psi}{A_{pp}\mu}]\delta\omega_q \!+\! g\delta\theta\,,\\
\tfrac{{\rm d}}{{\rm d}t}\delta\theta&=\sigma_\psi\delta\omega_q\,,\\
\tfrac{{\rm d}}{{\rm d}t}\delta\omega_q&=-\tfrac{\mu}{J_{qq}}[C_{qq}\delta\omega_q+B_{qn}\delta v_n]\,,
\end{align}
\label{eq:subsystem_theta}
\end{subequations}
\noindent is similar to Eq.~(\ref{eq:system2}). However, frequencies and bifurcations may differ depending on the numerical values of the resistance coefficients and upon the moments of inertia. This means that  there may be parameter regimes where $\theta$ decays without oscillations while $\psi$ oscillates, or vice versa.

Evaluating the stability exponents of Eqs.~(\ref{eq:subsystem_psi}) and (\ref{eq:subsystem_theta}) for the semi-circular fibres in Table~\ref{tab:coeffs} shows that the fixed point $\psi^*=0$ ($\sigma_\psi=1$) is stable in the Stokes limit. Both sub-systems predict oscillations for all semi-circular fibres from Table~1 in the main text, except for the fibres with $\kappa=100$ and $L=1$mm, which do not show oscillations at all. 

Moreover, the period time for the dynamics (\ref{eq:subsystem_theta}) is larger than that of the dynamics (\ref{eq:subsystem_psi}) by approximately a factor 2 for all cases showing oscillations.
The time scale of damping is significantly smaller than the period time, with the exception of the $\kappa=50$ and $L=2$mm semi-circular fibre.
In conclusion, oscillations in the experiments are expected to be slightly easier to be seen in $\psi(t)$, but only if the initial $\psi(t=0)$ is significantly different from zero (or if the initial $\omega_n(t=0)$ is different from zero). 

We note that if $B_{nq}$ and $B_{qn}$ have the same sign, both fixed points $\psi^*=0$ and $\psi^*=\pi$ are unstable. This is the case for taco-shaped disks
which continue to rotate as they settle~\cite{miara2024dynamics}, they do not align.
This follows from the analysis described above, by 
 solving $\ve F^{(0)}_{\rm h}+\ve F_g=0$ and $\ve T^{(0)}_{\rm h}=0$ for $\ve v^\ast $ and $\ve\omega^\ast $, just
 as in Ref.~\cite{huseby2024helical}. In this way one obtains  $\psi$ as a function $\theta$: 
\begin{align}
\label{eq:pp}
    \sin\psi\sin\theta =C_0\cos(\theta)^{-\tfrac{B_{nq}(A_{nn}C_{qq}-B_{qn}^2)}{B_{qn}(A_{qq} C_{nn}-B_{nq}^2)}}
\end{align}
with constant of motion $C_0$ (the full solution is found  from (\ref{eq:pp}) by reflections).  The resulting phase portrait forms bands of closed orbits around center fixed-points separated by manifolds connecting saddle points at $\psi^*=m\pi$ and $\theta^*=n\pi$ with integer values of $m$ and $n$.
 This explains Fig.~5 in Ref. \cite{miara2024dynamics} which shows one half the phase portrait. Such phase portraits were also obtained by numerical simulation  in Ref.~\cite{miara2024dynamics}. \citet{gonzalez2004dynamics} discuss the settling angular dynamics of asymmetric particles  in the Stokes limit solely in terms of mobility matrix $\mathbb{b} = (\ma B^{\sf T} -\ma A \ma B^{-1} \ma C)^{-1}$.
 This is consistent with Eq.~(\ref{eq:pp}), because it can be expressed entirely in terms
 of elements of $\mathbb b$: the exponent in Eq.~(\ref{eq:pp})
 is just $-b_{qn}/b_{nq}$. Our linear stability analysis shows, however, that
 the mobility matrix  $\mathbb b$ alone does not suffice to determine the approach
 of the steady state. 

\section{Slender-body theory to order $\mathrm{Re}_a$}
\label{app:slender}
\citet{Khayat_Cox_1989} computed the hydrodynamic force per unit length $\ve f_{\rm h}(s)$ 
on a segment of a slender body held fixed in a uniform flow $\ve u = u \hat{\ve u}$,
\begin{subequations}
\label{eq:kc}
\begin{align}
\label{eq:fs}
\ve f_{\rm h} &= 2\pi \rho_f \nu u a^2\Big\{(\ve t \ve t^{\sf T} - 2\ma I)\hat{\ve u} \frac{1}{\log \kappa}  + \Big[ (\ve t \ve t^{\sf T} - 2\ma I)\big( \ve J+ \hat{\ve u} \log2\epsilon\big) + \big (\tfrac{3}{2}\ve t \ve t^{\sf T} - \ma I\big)\hat{\ve u}\Big] \frac{1}{(\log\kappa)^2}+ \ldots\Big\}\,,\\
\label{eq:fullJdef}
\ve J(s) &= \frac{1}{4}\Big( \int_{-1}^{s-\epsilon}+ \int_{s+\epsilon}^1 \Big){\rm d}s' \ma G(\ve x-\ve x') (2\ma I -\ve t' {\ve t'}^{\sf T})\hat{\ve u}\,,\\
G_{ij}(\ve x)& = \delta_{ij} \Delta \Psi - \partial_i \partial_j \Psi (\ve x)\,,\quad\mbox{with}\quad \Psi(\ve x) = \frac{2}{ {\rm Re}_a} \int_0^{\xi_1}\!\!\!\!\!\!{\rm d}\xi\, \frac{1-{\rm e}^{-\xi}}{\xi} \quad\mbox{and}\quad\xi_1= \tfrac{1}{2} {\rm Re}_a (x-\hat {\ve u}\cdot \ve x)\,.
\end{align}
 \end{subequations} 
Here $\ve x=\ve x(s)$ and $\ve x'=\ve x(s')$ 
are locations along
the contour of the fibre, made non-dimensional by dividing by $a$, and parameterised by $s$
and $s'$. The corresponding tangent vectors are 
$\ve t =\ve t(s)$ and $\ve t' = \ve t(s')$. The infinitesimal positive $\epsilon$ is introduced because the integrand has a singularity
at $s=s'$ (a divergence for Stokes contribution, a finite jump for Re$_a$-contribution). 
\citet{Khayat_Cox_1989} used Eq.~(\ref{eq:kc}) to compute force and
torque on a straight fibre.

From $\ve f_{\rm h}(s)$, 
force and torque w.r.t. to the centre-of-mass $\ve x_{\rm com}$ are computed by line integrals
\begin{subequations}
\begin{align}
\label{eq:hydF}
\ve F_{\rm h}& = a\int_{-1}^1 \!{\rm d}s\, \ve f_{\rm h}(s)\,,\\
\ve T_{\rm h}& = a\int_{-1}^1 \!{\rm d}s\, [\ve x(s)-\ve x_{\rm com}] \wedge \ve f_{\rm h}(s)
\label{eq:tdef}
\end{align}
\label{eq:hydFT}
\end{subequations}
along the fibre. 
 To perform the integrals, we parameterise the curved planar fibres as
\begin{equation}
\label{eq:param}
\ve x(s) = \ve x_0+R\, \ve p \cos(\tfrac{\pi}{M}s)   + R \ve q\sin(\tfrac{\pi}{M}s)\,,
\end{equation}
where $R=M/\pi$ is the non-dimensional radius of curvature, $M=2$ for a semi-circle and $M=4$ for a quarter circle.
The centre-of-mass $\ve x_{{\rm com}}$  is determined by
\begin{equation}
\ve x_{{\rm com}} -\ve x_0 =\ve p\, \tfrac{R}{2} \int_{-1}^{1}\! \!\!{\rm d}s\, \cos(\tfrac{\pi}{M} )s = 
\ve p\, (\tfrac{M}{\pi})^2\sin(\tfrac{\pi}{M}) \,.
\end{equation} 
To compute the fluid-inertia
contribution to force and torque at small but non-zero Re$_a$,
we extract from Eqs.~(\ref{eq:kc})
the contributions linear in Re$_a$,
by a Re$_a$-expansion of $\Psi(\ve x)$. It follows that to order Re$_a$, the fluid-inertia contribution to force and torque  is given by
\begin{subequations}
\label{eq:kc2}
\begin{align}
\ve f_{\rm h}^{(1)} &= 2\pi \rho_f \nu u  (\ve t \ve t^{\sf T} - 2\ma I) \ve J^{(1)} /(\log\kappa)^2\,,\\
\label{eq:J1def} \ve J^{(1)} (s) &= 
\frac{1}{4}
\int_{-1}^1{\rm d}s' \ma G^{(1)}(\ve x-\ve x') (2\ma I -\ve t' {\ve t'}^{\sf T} )\hat{\ve u}\,,\\
G^{(1)} _{ij}(\ve x)& = \delta_{ij} \Delta \Psi^{(1)} - \partial_i \partial_j \Psi^{(1)} (\ve x)\,,\quad\mbox{with}\quad
\label{eq:psi1}
\Psi^{(1)} (\ve x) = -\tfrac{1}{8} {\rm Re}_a 
(\sqrt{|\ve x|^2}-\hat{\ve u}\cdot \ve x)^2\,.
\end{align}
 \end{subequations} 
To order Re$_a$, Eq.~(\ref{eq:kc2}) shows that the fluid inertia torque contains at most three powers of 
$\hat u_p$, $\hat u_q$, and $\hat u_n$. Evaluating the integrals (\ref{eq:J1def}) and (\ref{eq:tdef}), we obtain 
\begin{subequations} 
\label{eq:torque_re_u}
\begin{align}
 T_{p,M}^{(1)}/(2\pi \mu u a^2)& = 
\tfrac{ au}{\nu(\log \kappa)^2}  a_{p,M} {\hat u}_q {\hat u}_n\,,\\
T_{q,M}^{(1)}/(2\pi \mu u a^2) &=
 \tfrac{au}{\nu(\log \kappa)^2}  a_{q,M} {\hat u}_p {\hat u}_n\,,\\
T_{n,M}^{(1)}/(2\pi \mu u a^2) &=
\tfrac{ au}{\nu(\log \kappa)^2}
\big[a_{n,M}{\hat u}_q  {\hat u}_p
-b_{n,M} {\hat u}_q -c_{n,M} \hat u_q 
\hat u_n^2- d_{n,M} \hat u_q^3\big]\,.
\end{align}
 \end{subequations}
This is Eq.~(4) in the main text, with $\ve v = -\ve u$. As explained there, $M$ parameterises the  central angle $2\pi/M$ of the circle segment.
For the coefficients we find:
 \begin{subequations}
 \label{eq:coeffs}
 \begin{align}
a_{p,M}&\!=\!-\tfrac{M^2}{36 \pi^3}\!\big[84\pi\!-\!M(57s_{1,M}\!+\!12s_{2,M}\!+\!s_{3,M})\big],\\
a_{q,M}&=\tfrac{M^2}{36\pi^4}\big[84\pi^2+3M^2(-2+c_{1,M}+2c_{2,M}-c_{3,M})-M\pi(105s_{1,M}- 12s_{2,M}+s_{3,M})\big]\,,\\
a_{n,M}&=\tfrac{M^3}{48\pi^4}s_{1,M}\big[12\pi(1-c_{2,M})+M(16s_{1,M}-6s_{2,M}-s_{4,M})\big]\,,\\
b_{n,M}&\!=\!\tfrac{M^2}{32\pi^4}s_{1,M}\big[12\pi^2+\!M^2(1\!-\!c_{4,M})\!-\!10M\pi s_{2,M}\big]\,,\\
c_{n,M}&=-\tfrac{M^2}{64\pi^4}s_{1,M}(2\pi-Ms_{2,M})^2\,,\\
d_{n,M}&=\tfrac{M^3}{32\pi^4}s_{1,M}s_{2,M}(2\pi-Ms_{2,M})\,,
\end{align}
\end{subequations}
where $c_{n,M}=\cos(n\pi/M)$ and $s_{n,M}=\sin(n\pi/M)$. Evaluating $\ve T^{(1)}$ for large values of $M$ using Eqs.~(\ref{eq:torque_re_u}) and (\ref{eq:coeffs}), we find that  $T_q^{(1)}\to 0$
while $T_p^{(1)}$ and $T_n^{(1)}$ combine to the torque on a slender straight wire, 
$\ve T^{(1)}\!=\!-\!\tfrac{5 \pi}{3} \rho_{\rm f} a^3  (\ve q\cdot\ve v) (\ve q\wedge\ve v) /(\log\kappa)^2$, 
where $\ve q$ is the vector along the symmetry axis of the straight wire. 
This result is  consistent with Eqs.~(6.19) and (6.22) of \citet{Khayat_Cox_1989}
for a straight fibre. 
For a ring ($M=1$) we find $\ve T^{(1)}=14a^3 (\ve n\cdot\ve v) (\ve n\times\ve v)\rho_{\rm f}/(3 \pi(\log\kappa)^2)$. Here $\ve n$ is the unit vector 
along the particle symmetry axis. The fluid-inertia torques for ring and rod
have opposite signs, as expected~\cite{Dabade_2015,Jiang_2021}.

\section{Slender-body theory to higher orders of  ${\rm Re}_a$}
\label{app:correlation}
\begin{figure}[b]
\begin{overpic}[width=1.cm]{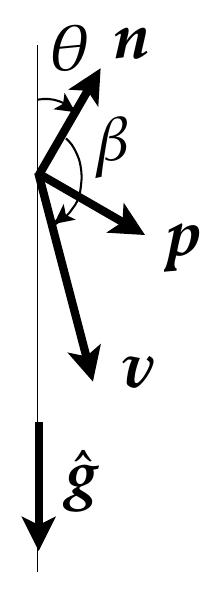}
\end{overpic}
\caption{\label{fig:betadef} {\em Parameterisation of inertial torque at larger Re$_a$.} Definition of the angle $\beta$ used to parameterise  fluid-inertia corrections to force and torque for larger Re$_a$. The vectors $\ve p$ and $\ve n$ are part of the particle-fixed coordinate system [Fig.~1({\bf a}) in the main text], $\theta$ is the tilt angle, $\ve v$ is the settling velocity, and  $\hat {\ve g}$ is the direction of gravity. The four vectors $\ve p$, $\ve n$,  $\ve v$, and  
$\hat {\ve g}$  lie in the same plane.}
\end{figure}

To compare with experiments, we must deal with
 the fact that Re$_a$ is not small, so that
 (\ref{eq:torque_re_u}) and the corresponding corrections for the force do not directly apply.
The standard procedure is to account for higher Re$_a$ by parameterisations of the fluid-inertia corrections.
For particles with axisymmetry and fore-aft symmetry this is straightforward for Re$_a$ up to the order of ten~\cite{Jiang_2021,roy2023orientation,bhowmick2024inertia}. 

For our slender curved fibres, we can use the full slender-body theory from Ref.~\cite{Khayat_Cox_1989} to evaluate fluid-inertia contributions to torque and force for larger Re$_a$.
To this end, we subtract the Stokes contribution (Re$_a=0$)
from Eqs.~(\ref{eq:kc}) to obtain a set of equations on the same form as Eqs.~(\ref{eq:kc2}), but with $\Psi^{(1)}(\ve x)$ replaced by
the fluid-inertia contribution
\begin{align}
\Psi^{({\rm i})}(\ve x) &= \frac{2}{ {\rm Re}_a} \int_0^{\xi_1}\!\!\!\!\!\!{\rm d}\xi\,\Big( \frac{1-{\rm e}^{-\xi}}{\xi}-1\Big)
\quad \mbox{with}\quad \xi_1= \tfrac{1}{2} {\rm Re}_a (x-\hat {\ve u}\cdot \ve x)\,.
\end{align}
The resulting set of equations yields the fluid inertia forces and torques for our fibres for any Re$_a$ within the the slender-body approximation 
of \citet{Khayat_Cox_1989}. 
Since we could not evaluate the resulting integrals  [Eqs.~(\ref{eq:fullJdef}) and (\ref{eq:hydFT})]
in closed form, we resorted to numerical integration. 
In this we obtained the fluid-inertia corrections to force and torque in the particle-fixed basis as functions of Re$_a$ and the orientation of $\hat{\ve u}$.
We checked that the numerical results simplify to Eq.~(4) in the main text in the limit of small Re$_a$. 

Our numerical results show that the analysis performed in the limit of small Reynolds numbers in the main text is qualitatively unchanged, but there are quantitative differences:
the steady state remains at $\ve\omega^*=0$, $\psi^*=0$, and $v_q^*=0$, but the values of $\theta^*$, $v_p^*$, and $v_n^*$ obtain corrections due to the force contribution from fluid inertia, and because Re$_a$ is not infinitesimal. Moreover, the location of the pitchfork bifurcation changes, it is no longer given by Eq.~(6) in the main text.
We determine $\theta^*$, $v_p^*$, and $v_n^*$ by evaluating the fluid-inertia corrections $F^{(\rm i)}_p({\rm Re}_a,{\beta})$, $F^{(\rm i)}_n({\rm Re}_a,{\beta})$ and $T^{(\rm i)}_q({\rm Re}_a,{\beta})$ with $\ve\omega^*=\psi^*=v_q^*=0$ (the other force and torque components are zero at this point) numerically for a large number of values of ${\rm Re}_a$ and $\beta$. Here $\beta$ is the angle between $\ve n$ and the velocity $\ve v$
(see Fig.~\ref{fig:betadef}), and Re$_a = av^\ast/\nu$ is the particle Reynolds number defined in the main text.

The steady states are obtained by finding the values ${\rm Re}_a^*$, $\beta^*$ and $\theta^*$ that solve the system
\begin{subequations}
\begin{align}
F_p &= -\mu A_{pp} v_p + F^{(\rm i)}_p({\rm Re}_a,\beta) + mg\sin\theta=0\,,\\
F_n &= -\mu A_{nn} v_n + F^{(\rm i)}_n({\rm Re}_a,\beta) - mg\cos\theta=0\,,\\
T_q &= -\mu B_{qn} v_n + T^{(\rm i)}_q({\rm Re}_a,\beta)=0\,,
\end{align}
\label{eq:ForceTorqueInertiaCondition}
\end{subequations}
where $v_n=v\cos\beta$, $v_p=v\sqrt{1-\cos^2\beta}$ and $v={\rm Re}_a\nu/a$.
Eq.~(\ref{eq:ForceTorqueInertiaCondition}) has one fixed point with $\theta^*=\pi/2$ and $v_n^*=0$ as in the Stokes limit, but $v_p^*$, obtained by solving $-\mu A_{pp} v_p + F^{(\rm i)}_p({\rm Re}_a,\pi/2) + mg=0$, takes lower values due to the inertial correction $F^{(\rm i)}_p$.

This fixed point becomes unstable in a pitchfork bifurcation, as described in the main text. 
The location of the bifurcation is determined by first solving $\lim_{v_n\to 0}T_q/v_n$ for $v_p$, and then inserting the resulting solution together with $\theta=\beta=\pi/2$ into the expression for $F_p$.
The bifurcation occurs at parameter values where $F_p=0$. The sign of $F_p$ for a given set of parameters tells whether the system is before ($F_p<0$) or after ($F_p>0$) the bifurcation.
After the bifurcation, we numerically solve 
Eqs.~(\ref{eq:ForceTorqueInertiaCondition}) to obtain the stable fixed points $\theta^*$ and $v_g^*=-v_n^*\cos\theta^*+v_p^*\sin\theta^*$ shown in Fig.~3 in the main text.

\section{Data analysis}
\label{app:analysis}

\citet{tatsii2024shape} measured the settling speeds of straight and curved slender fibres in quiescent air using the G\"ottingen turret \cite{bhowmick2024inertia}. 
Here we extracted fibre orientation from the video recordings of \citet{tatsii2024shape},  to determine how shape-symmetry breaking affects the angular dynamcis of the settling fibres.
The G\"ottingen turret has two pairs of cameras, top and bottom. The two cameras in a pair are orthogonal to each other
in a plane perpendicular to the direction of gravity. Details of the setup are described by \citet{bhowmick2024inertia}.
The fibers used in our experiments were precisely created using a 3D printer, specifically the Photonic Professional GT model by Nanoscribe GmbH (2015), capable of achieving a printing precision of $<1\mu$m.
The fibres were printed by polymerising acrylic resin $\text{IP-S }(\text{CH}_{1.27}\text{N}_{0.086}0_{0.37})$ with density of 1200 kg/m$^3$. 
The uniform polymerisation of the resin during printing and subsequent curing, combined with the fact that the printer always produces a solid body without voids, minimises variations in particle density. 
The 3D printer requires the 3D CAD models of the fibres to produce the fibers.
To check for possible defects that could lead to deviations in fibre size and shape, we examined all fibres under a microscope after printing and the defected ones are removed from the batch as explained in \cite{tatsii2024shape}. 
All remaining ones were found to match the expected geometric specifications. 
See Figure S1 in the supplementary material by \citet{tatsii2024shape} for example photos and 3D scan profiles of particles taken after printing.

We analysed all experiments where the fibre was visible and in focus in at least in one camera pair (top and/or bottom), in order to obtain reliable estimates for the fibre orientation. In their analysis of the setting speed, \citet{tatsii2024shape}
also considered experiments where the fibre is not captured by all cameras. Therefore the number $N$ of experiments analysed here is smaller than in Table~1 in Ref.~\cite{tatsii2024shape}.

Fibre orientations were tracked  in a three-stage process: firstly,  experimental images are pre-processed; secondly, 
synthetic projection images of
fibres with different orientations were created using 
their 3D CAD models; and finally, fibre orientations were extracted by matching  experimental and synthetic projection images.
During pre-processing, all frames corresponding to a given fibre were cropped to a uniform size. This guarantees that the entire projection  of the fibre
is contained in the frame, regardless of fibre orientation. Each fibre image was then shifted so that the left edge of the frame such the top left corner of the rectangular box bounding the particle image was in the same pixel for all images. A fixed threshold based on the entire camera field of view was used for binarization of the particle images. However, it was found that shifting all thresholded images to the left triggered greater sensitivity in finding the best match for quarter circles and some of the larger semi-circles (specifically hc 2000x40 E06 and hc 2000x20 E08). This was attributed to the bounding box being wider than the rest for most of these fibres and/or various possible experimental uncertainties (as outlined below).  As a result, we re-analysed data obtained for the these cases by determining a threshold for each cropped image and adjusting the matching library database to contain a finer angular resolution (see below).

To create synthetic images, we used the 3D-CAD model employed for printing the fibres. 
we varied the Euler angles of the fibre 3D model, with resolution $\delta \phi = 0.02\pi, \delta \theta = 0.01\pi, \delta\psi = 0.02\pi$ for most semi-circles and  $\delta \phi = 0.02\pi, \delta \theta = 0.005\pi, \delta\psi 0.005\pi$ for quarter circles and some semi-circles. 
We projected the 3D model to the two orthogonal camera planes such that the size of the largest dimension of the produced image (in pixels) was $L/r$, where $L$ is the length of the largest dimension of the fibre in $\mu$m and $r$ is the effective camera resolution in $\mu$m per pixel, i.e. 6.75 $\mu$m/pixel. This ensures that the size of synthetic images matches those captured by the cameras. 
The best match between experimental and synthetic image was selected to be the one with the smallest pixel-wise distance summed over views of the camera pair.
 
 The shape symmetry of the fibres implies that
there are four equivalent triplets of Euler angles for a given fibre orientation (Section~\ref{app:euler}).  To resolve this ambiguity in Figs.~\ref{fig:hc_2000x50} and~\ref{fig:hc_1000x50}, we chose time series of Euler angles that yield continuous smooth trajectories. For the other experimental data shown in Appendix~\ref{app:data}, we constrained 
 the angles as follows: $\theta$ to the interval $[0,\pi/2]$, and $\psi$ to $[-\pi/2,\pi/2]$.

Our analysis is subject to the following errors.
In some cases, out-of-focus images give
rise to distorted projections, so that maximising pixel-wise overlap may cause systematic errors. 
The size of the experimental fibre projection may deviate from the expected dimension for a number of reasons. These include changes in magnification along the camera axis, slight blurring of the particle image, and imprecise thresholding. Quantifying these errors is challenging, as they are likely to vary independently across different experiments. However, based on the standard deviation values reported in 
Table 1 in the main text, their impact on the measured angles appears to be small.
Camera misalignment can be ruled out as a source of error as three-dimensional camera calibrations revealed that the angles between the two cameras in a pair deviate by less than 1 degree from a perfect right angle.
  The temperature in the room is maintained at about \SI{22.4}{\degreeCelsius} during the experiments. To estimate the maximum possible change in air temperature in the settling chamber, we observed an increase of about 1 degree when the LED runs continuously for 15 minutes (normally it runs for a few seconds every few minutes). This conservative estimate of the temperature increase can change the viscosity of the air by a maximum of 1\%. 
  
  It should be noted that the initial conditions of the particles injected into the settling chamber, including the initial release velocity, angular velocity and orientation, are not the same between the different experimental runs. 
 Variations in transient dynamics can occur since the images are taken at a fixed distance from the injection site, even if the particles have the same shape, see Fig.~\ref{fig:hc_1000x50}. 
 The differences observed in cases that appear to have reached a steady state, are likely due to other experimental uncertainties. However, the low scatter of the experimental results, as indicated by the low value of the standard deviation, suggests that they are highly reliable.

\section{Summary of experimental results}
\label{app:data}
\begin{figure*}[p]
\begin{overpic}[width=0.9\columnwidth]{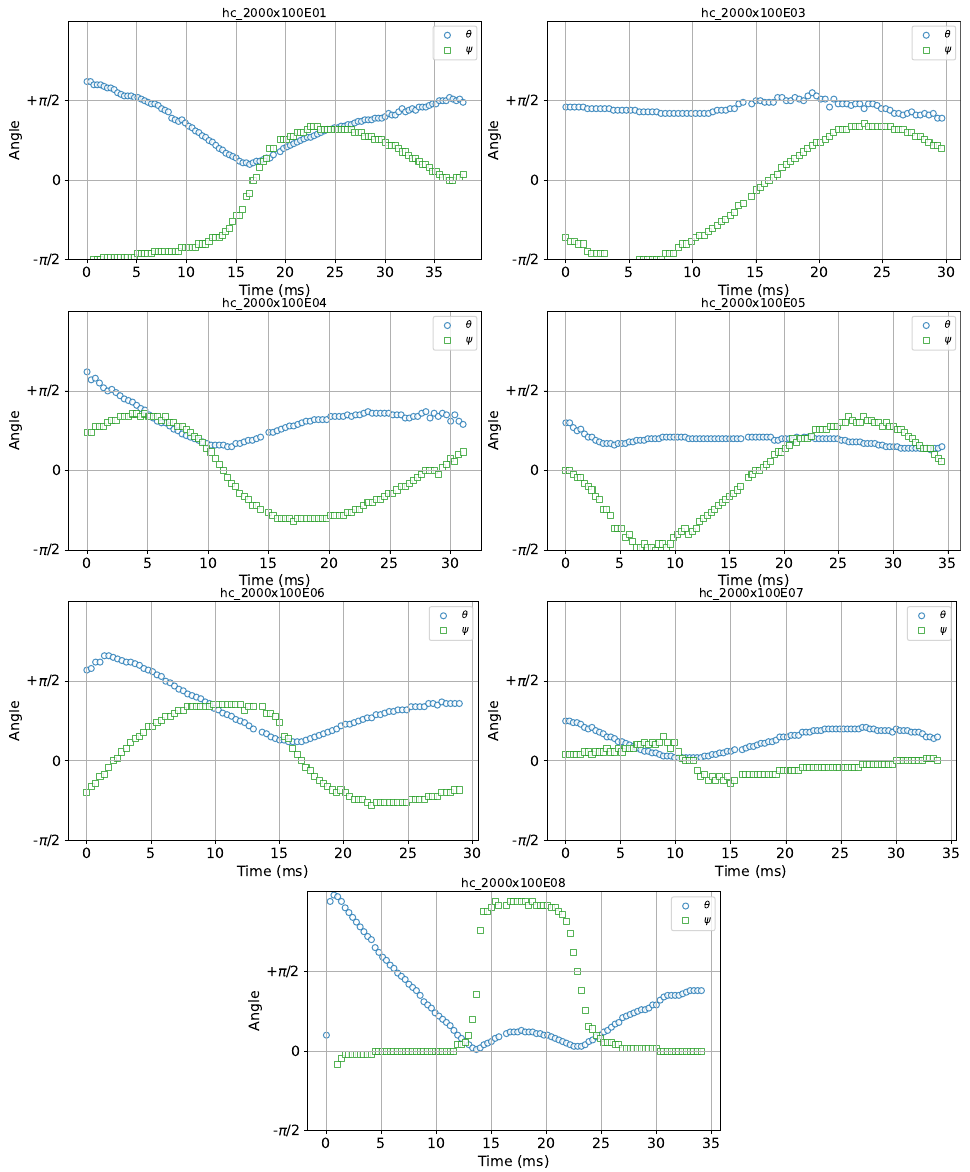}\end{overpic}
\caption{\label{fig:si} Semi-circular fibres with aspect ratio $\kappa = 20$, length $L=2000\,\mu$m, and  diameter  $d = 100\,\mu$m. Shown are the angles $\theta$ and $\psi$ from upper and lower cameras. }
\label{fig:hc_2000x50}
\end{figure*}

Figures~\ref{fig:si} and \ref{fig:sf} show the angles
extracted from video recordings 
of Ref.~\cite{tatsii2024shape} as described in Appendix \ref{app:analysis}.
Shown are results for two different semi-circular fibres. Figures~\ref{fig:si2} to \ref{fig:sf2} show snapshots of the projections of steadily settling fibres in onto the camera planes of the bottom cameras.


%

\begin{figure*}[p]
\begin{overpic}[width=0.9\columnwidth]{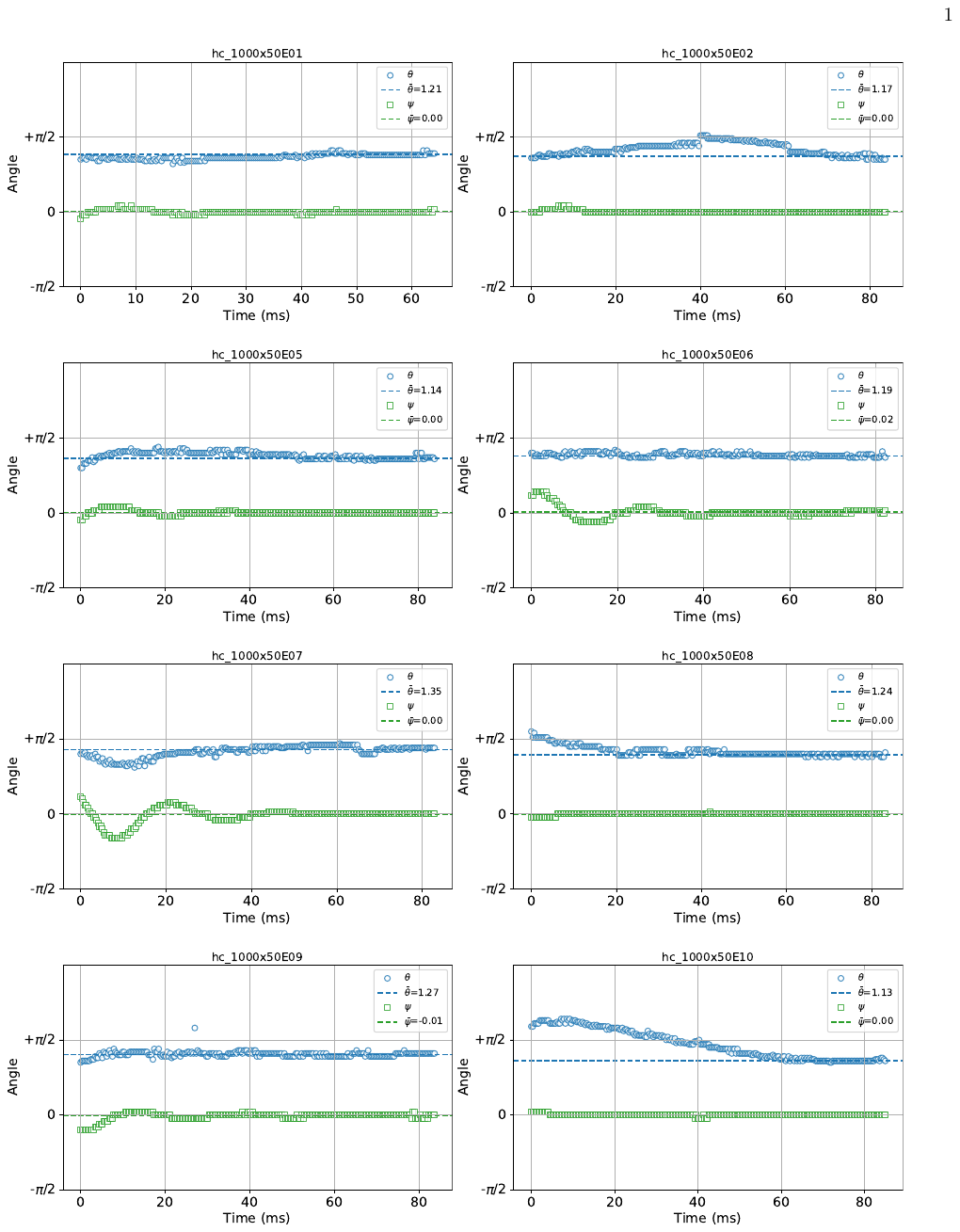}\end{overpic}
\caption{\label{fig:sf} Semi-circular fibres with with aspect ratio $\kappa = 20$, length $L=1000\,\mu$m, and  diameter  $d = 50\,\mu$m. Shown are the angles $\theta$ and $\psi$ from upper and lower cameras, as well as the average angles (averaged over the frames from the bottom cameras). 
\label{fig:hc_1000x50}}
\end{figure*}

\begin{figure*}[p]
\begin{overpic}[width=0.9\columnwidth]{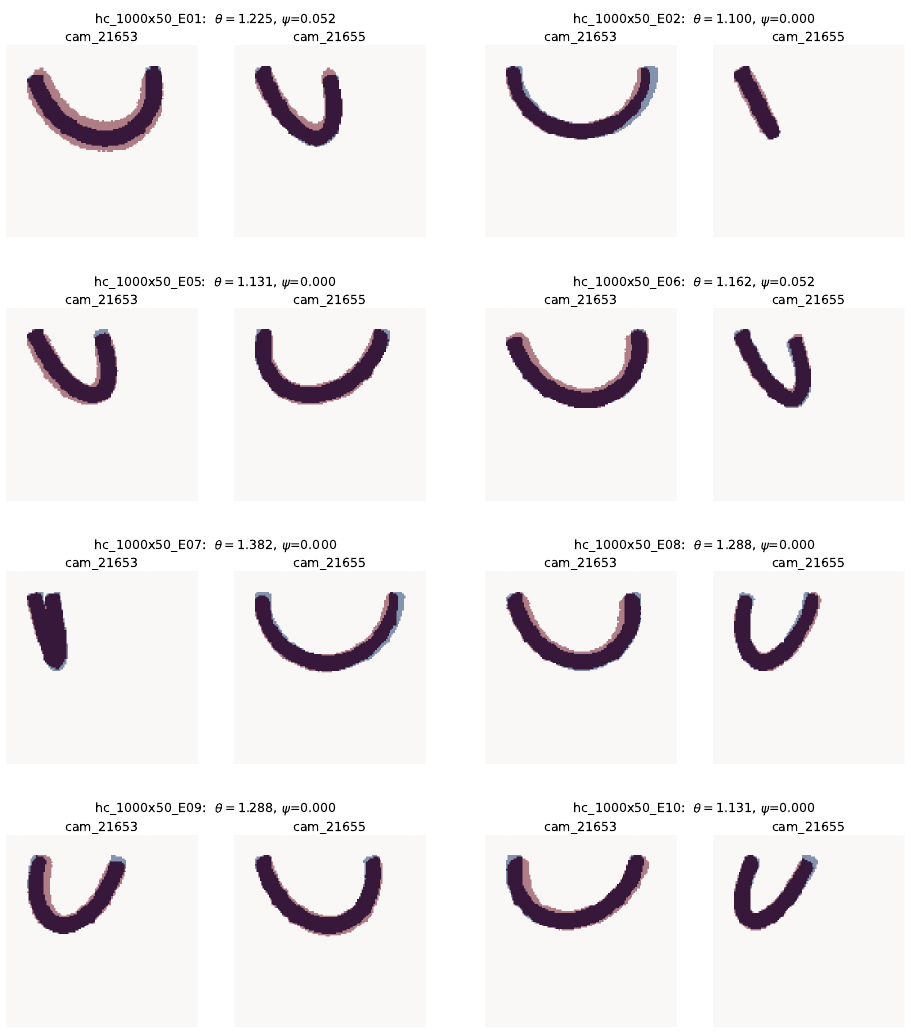}\end{overpic}
\caption{\label{fig:si2} Snapshots of steady state for semi-circular fibres with with aspect ratio $\kappa = 20$, length $L=1000\,\mu$m, and  diameter  $d = 50\,\mu$m. 
Shown are the fibres observed in the planes of the bottom cameras towards the end of experimental run (red), as well as the extracted orientation (blue).}
\end{figure*}

\begin{figure*}[p]
\begin{overpic}[width=0.43\columnwidth]{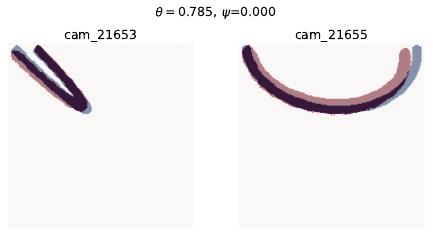}\end{overpic}
\caption{Same, for semi-circular fibre with aspect ratio $\kappa = 50$,  length $L=2000\,\mu$m,  and  diameter  $d = 40\,\mu$m. }
\label{fig:hc_2000x40}
\end{figure*}

\begin{figure*}[p]
\begin{overpic}[width=0.43\columnwidth]{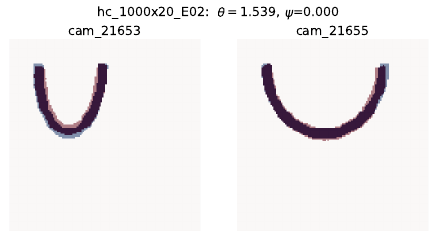}\end{overpic}
\caption{Same, for semi-circular fibre with aspect ratio $\kappa = 50$,   length $L=1000\,\mu$m,  and  diameter  $d = 20\,\mu$m. }
\end{figure*}

\begin{figure*}[p]
\begin{overpic}[width=0.9\columnwidth]{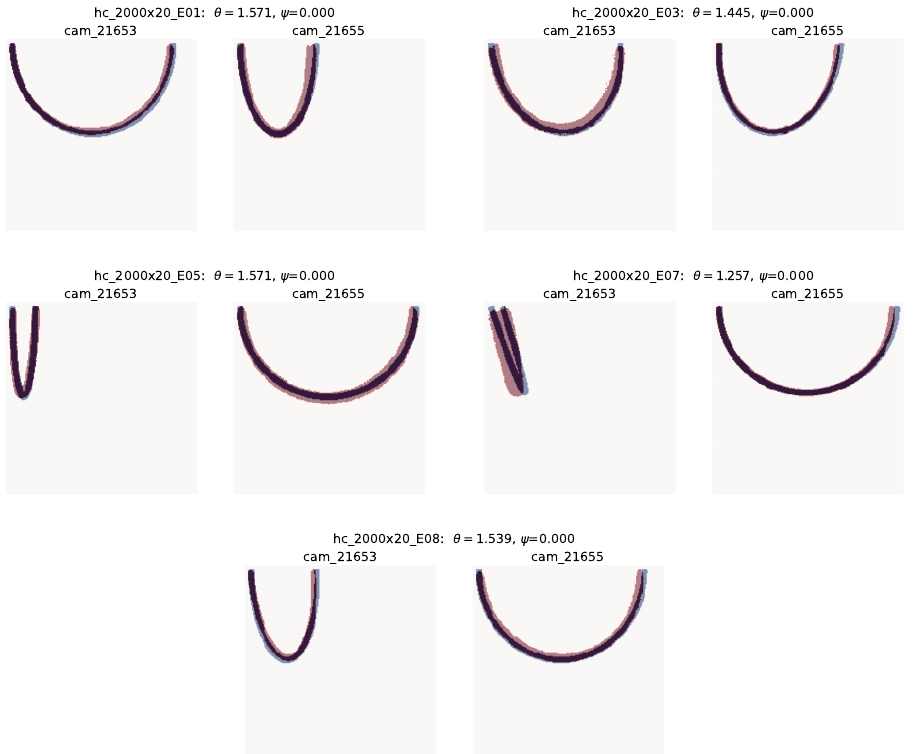}\end{overpic}
\caption{Same, for semi-circular fibres with aspect ratio $\kappa = 100$, length $L=2000\,\mu$m,  and  diameter  $d = 20\,\mu$m.}
\end{figure*}

\begin{figure*}[p]
\begin{overpic}[width=0.9\columnwidth]{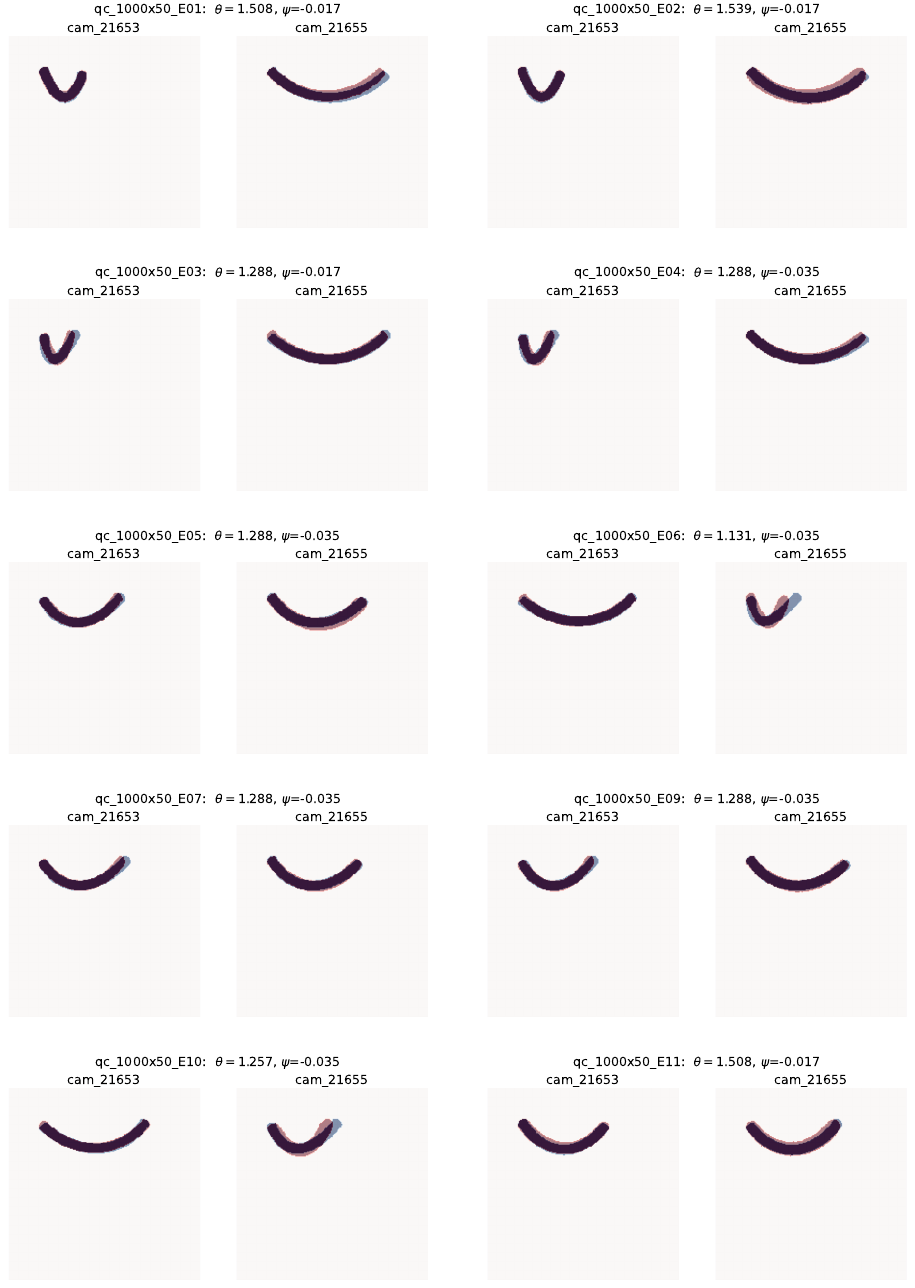}\end{overpic}
\caption{ Same, for quarter-circular fibres  with aspect ratio $\kappa = 20$, length $L=1000\,\mu$m,  and  diameter  $d = 50\,\mu$m.}
\end{figure*}

\begin{figure*}[p]
\begin{overpic}[width=0.43\columnwidth]{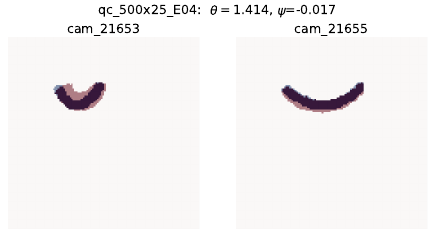}\end{overpic}
\caption{Same, for quarter-circular fibre  with aspect ratio $\kappa = 20$, length $L=500\,\mu$m,  and  diameter  $d = 25\,\mu$m.}
\end{figure*}

\begin{figure*}[p]
\begin{overpic}[width=0.9\columnwidth]{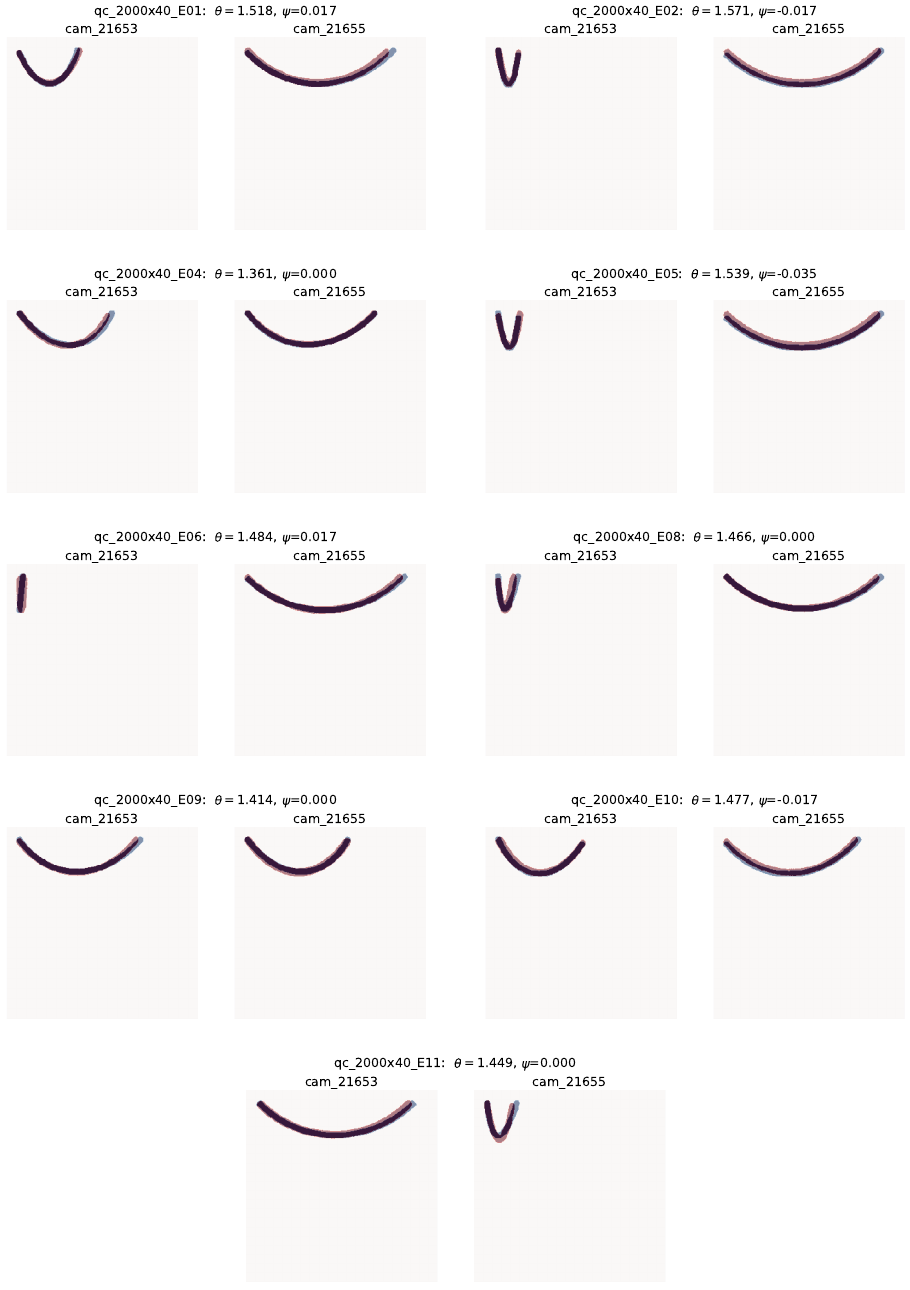}\end{overpic}
\caption{Same, for quarter-circular fibres with  aspect ratio $\kappa = 50$, length $L=2000\,\mu$m,  and  diameter  $d = 40\,\mu$m.}
\end{figure*}

\begin{figure*}[p]
\begin{overpic}[width=0.9\columnwidth]{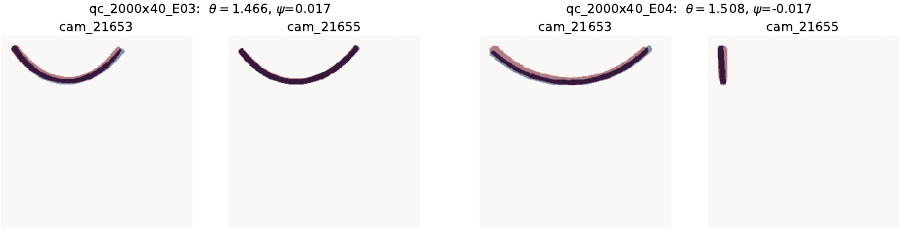}\end{overpic}
\caption{\label{fig:qc0} Fig.~S10 continued. 
}
\end{figure*}

\begin{figure*}[p]
\begin{overpic}[width=0.9\columnwidth]{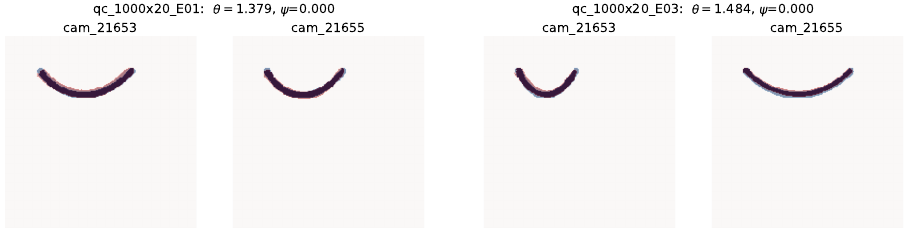}\end{overpic}
\caption{Same, for quarter-circular fibre with aspect ratio $\kappa = 50$, length $L=1000\,\mu$m,  and  diameter  $d = 20\,\mu$m.}
\end{figure*}

\begin{figure*}[p]
\begin{overpic}[width=0.43\columnwidth]{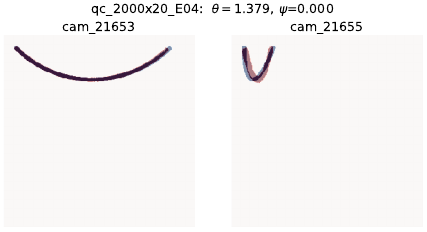}\end{overpic}
\caption{\label{fig:sf2} Same, for quarter-circular fibre with  aspect ratio $\kappa = 100$, length $L=2000\,\mu$m,  and  diameter  $d = 20\,\mu$m.}
\end{figure*}
\vfill\eject